\documentclass[journal]{IEEEtran}
\usepackage{amsmath,amsthm}
\usepackage{amssymb}
\usepackage{amsfonts}
\usepackage[ruled]{algorithm2e}
\usepackage{graphicx}
\usepackage{float} 
\usepackage{subfigure}
\usepackage{tikz,tikz-3dplot}
\usepackage{tikzscale}
\usepackage{xcolor}
\usepackage{bm}
\usepackage{dsfont}
\usepackage{stfloats}
\usepackage[flushleft]{threeparttable}
\usepackage[numbers,sort &compress]{natbib} %

\usepackage[switch]{lineno}
\usepackage{mathrsfs}
\usepackage{bbm}

\DeclareMathOperator*{\argmin}{arg\,min}
\usepackage{svg}
\usepackage[colorlinks]{hyperref}
\hypersetup{colorlinks,breaklinks,linkcolor=blue,urlcolor=blue,anchorcolor=blue,citecolor=blue}
\usepackage{booktabs}
\usepackage{tabularx}
\usepackage{multirow}
\usepackage{lscape}

\usepackage{epstopdf}
\usepackage{booktabs}
\usepackage{caption}
\usepackage{enumitem}
\usepackage[flushleft]{threeparttable}

\usepackage{float}
\usepackage{microtype}
\usepackage{placeins}
\usepackage{xcolor}
\definecolor{deepblue}{rgb}{0,0,255}

\usepackage[defaultcolor=blue]{changes}

\usepackage{threeparttable}
\usepackage{dcolumn}
\newcolumntype{d}[1]{D{.}{.}{#1}}

\newcommand{\mat}[1]{\boldsymbol{#1}}

\definecolor{myblue}{rgb}{0,0,255}

\definecolor{myred}{rgb}{255,0,0}

\ifCLASSINFOpdf
\else
\fi

\usepackage{xcolor}
\definecolor{darkblue}{rgb}{0.0,0.5,0.5}

\begin{document}

\title{Efficient and Robust Freeway Traffic Speed Estimation under Oblique Grid using Vehicle Trajectory Data}

\author{\IEEEauthorblockN{Yang He, Chengchuan An, Yuheng Jia, \IEEEmembership{Member,~IEEE,} Jiachao Liu, Zhenbo Lu, and Jingxin Xia}

\thanks{This work was supported in part by the National Natural Science Foundation of China under Grants 52272309, 52202398, and 62106044, in part by the International Science and Technology Cooperation Project of Jiangsu Province under Grant BZ2023015, and in part by Natural Science Foundation of Jiangsu Province under Grant BK20210221 (Corresponding authors: Jingxin Xia).}
\thanks{Yang He, Chengchuan An, Zhenbo Lu, and Jingxin Xia are with the Intelligent Transportation System Research Center, Southeast University, Nanjing, 211189, China (e-mail: yanghe@seu.edu.cn, ccan@seu.edu.cn, luzhenbo@seu.edu.cn, xiajingxin@seu.edu.cn). }
\thanks{Yuheng Jia is with the School of Computer Science and Engineering, Southeast
University, Nanjing, 211189, China (e-mail: yhjia@seu.edu.cn). }
\thanks{Jiachao Liu is with the  Department of Civil and Environmental Engineering, Carnegie Mellon University, Pittsburgh, PA 15213, USA (e-mail: jiachaol@andrew.cmu.edu).}	}


\maketitle

\begin{abstract}

  Accurately estimating spatiotemporal traffic states on freeways is a significant challenge due to limited sensor deployment and potential data corruption.
  In this study, we propose an efficient and robust low-rank model for precise spatiotemporal traffic speed state estimation (TSE) using low-penetration vehicle trajectory data. 
  Leveraging traffic wave priors, an oblique grid-based matrix is first designed to transform the inherent dependencies of spatiotemporal traffic states into the algebraic low-rankness of a matrix. Then, with the enhanced traffic state low-rankness in the oblique matrix, a low-rank matrix completion method is tailored to explicitly capture spatiotemporal traffic propagation characteristics and precisely reconstruct traffic states.
  In addition, an anomaly-tolerant module based on a sparse matrix is developed to accommodate corrupted data input and thereby improve the TSE model robustness.
  Notably, driven by the understanding of traffic waves, the computational complexity of the proposed efficient method is only correlated with the problem size itself, not with dataset size and hyperparameter selection prevalent in existing studies.
  Extensive experiments demonstrate the effectiveness, robustness, and efficiency of the proposed model. 
  The performance of the proposed method achieves up to a 12$\%$ improvement in Root Mean Squared Error (RMSE) in the TSE scenarios and an 18$\%$  improvement in RMSE in the robust TSE scenarios, and it runs more than 20 times faster than the state-of-the-art (SOTA) methods.

  

\end{abstract}

\begin{IEEEkeywords}
Traffic state estimation, kinematic wave theory, low-rank representation, vehicle trajectory data.
\end{IEEEkeywords}



\IEEEdisplaynontitleabstractindextext

\IEEEpeerreviewmaketitle


\section{Introduction}

\begin{figure*}[t]
  \begin{center}
\includegraphics[width=6.5in]{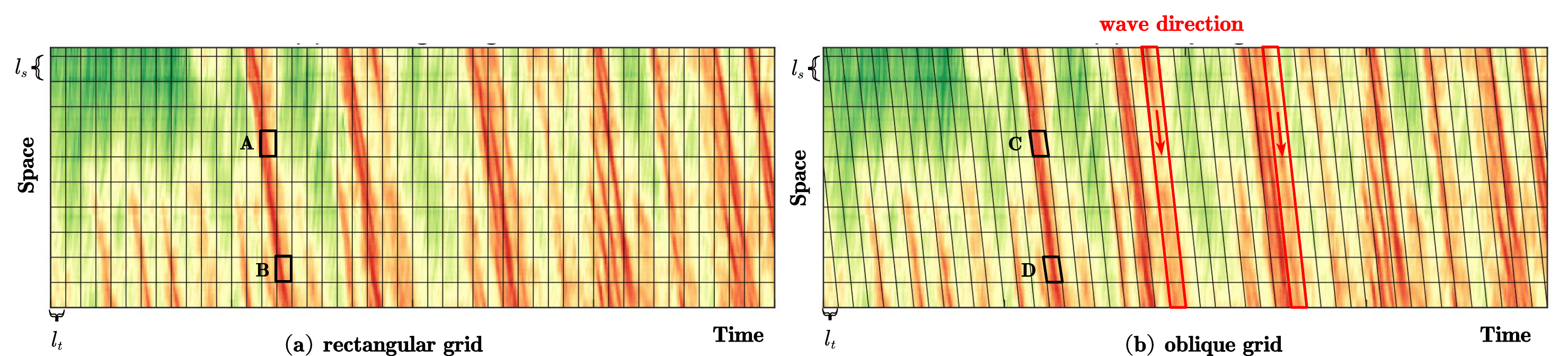}
\caption{Visualization of constructing a traffic state matrix (TSM). Traffic states exhibit high correlations along the direction of backward traffic waves. Conventional rectangular grid-based modeling in (a) is less desirable to effectively capture such correlations, as it simply vertically and horizontally divides the spatiotemporal region (e.g., cells A and B). In this study, we adopted the oblique grid-based modeling in (b), strategically positioning traffic state observations along the traffic wave direction into the same matrix column (e.g., cells C and D). This approach adeptly transforms the correlation of traffic states into the algebraic low-rankness of the matrix, therefore ensuring a low-rank representation method to proficiently capture the spatiotemporal correlations inherent in traffic states.}
\label{fig: Illustration}
  \end{center}
\end{figure*}


\subsection{Motivation}
Precise and complete traffic states (e.g., 5-sec traffic speed) provide reliable support for freeway proactive traffic control and management, especially in current and future connected and automated vehicular environments, e.g., connected and automated vehicle (CAV) cruise control, eco-driving, and dynamic routing planning \cite{boriboonsomsin2012eco,ozatay2014cloud, guanetti2018control}.
In practice, field traffic state measurements are often limited and noisy \cite{seo2017traffic, yang2022generalized, wang2023low}.
Fixed detectors are costly and often sparsely installed along the road, resulting in limited spatial coverage. Mobile sensors, benefiting from the advancements of connected vehicle (CV) technologies, provide more extensive spatial coverage. However, they suffer from sparsity in the temporal domain \cite{thodi2022incorporating} due to the low penetration rate in the current mixed conventional and connected environment. 
Reconstructing accurate traffic states on the freeway from the sparse and corrupted observations is still a challenging task that needs to be addressed in current applications of Intelligent Transportation Systems (ITSs).

\subsection{State-of-the-Art (SOTA)}

Initially, researchers carefully abstracted physical traffic flow characteristics and utilized traffic flow models including the first-order model like the well-known Lighthill-Whitham-Richards (LWR) to estimate traffic states \cite{yuan2012real,yuan2014NetworkWidea, work2010Traffic, wang2016Efficient, wang2016Multiple, duret2017traffic}, employing various data assimilation techniques. 
To more accurately capture complex traffic phenomena, higher-order models such as the Payne-Whitham (PW) models \cite{nanthawichit2003Application,liu2018Progressive},  Aw-Rascle-Zhang (ARZ) models \cite{seo2017traffic, wang2017Comparing}, and METANET models \cite{wang2022Realtime, zhao2022Generic, makridis2023Adaptive} have also been explored in TSE. 
An alternative approach to TSE assumes that the average speed of regular vehicles equals that of CVs \cite{bekiaris-liberis2016Highway, roncoli2016Use, fountoulakis2017Highway, bekiaris-liberis2017Highway, papadopoulou2018Microscopic}. This speed-uniformity assumption simplifies TSE by using a data-driven conservation equation model with Kalman filters \cite{zhao2022Generic}.  
Recent overviews of freeway TSE highlight these developments \cite{seo2017traffic, wang2022Realtime}. 
Benefiting from domain knowledge, these methods are physically interpretable and require a small amount of data. 
Despite the simplicity, model-based methods can be constrained by the capacity of the traffic flow models and assumptions made in the data assimilation process \cite{yang2022generalized}. Moreover, model-based methods usually require time-consuming and labor-intensive parameter calibration processes.

With the rapid progress in computation ability and wide availability of multi-source data, data-driven methods have flourished in TSE. The main approach of this category is to exploit the spatiotemporal dependencies from traffic data using various learning frameworks, such as adaptive smoothing kernel \cite{treiber2011reconstructing, chen2018adaptive,yang2022generalized, rempe2017phase}, Gaussian process \cite{yuan2021macroscopic,wu2023traffic}, deep learning \cite{thodi2022incorporating, rempe2022estimation, lu2020lane, wu2021inductive, nie2023towards, zhang2022tsr,xu2020ge}, low-rank matrix/tensor completion \cite{yu2020urban, wang2023low, nie2022correlating}, etc.
The most prevalent modeling approach is discretizing the spatiotemporal domain into a spatiotemporal grid/matrix/diagram as shown in Fig. \ref{fig: Illustration}(a). Then, fixed or mobile data are aggregated and transformed into partial observations of the grid.
The grid-based TSE modeling has become a popular framework due to its easy implementation and convenience in capturing high-dimensional spatiotemporal traffic flow dependencies \cite{rempe2022estimation, wang2023low}.

By decomposing the spatiotemporal domains into small unified grids,  \citet{rempe2022estimation} developed a convolutional neural network (CNN) to learn and reconstruct the spatiotemporal traffic speeds within these grids.  \citet{thodi2022incorporating} further incorporated kinematic wave priors into CNN by designing anisotropic kernels to capture directional traffic propagation characteristics. In addition, graph neural networks \cite{wu2021inductive, nie2023towards} and generative adversarial networks \cite{zhang2022tsr,xu2020ge} are also applied.
However, these deep learning-based methods may require massive and high-quality training data. 
It is worth noting that obtaining a suitable training dataset may not always be feasible in practice \cite{wu2023traffic}. Although the training data can be collected from traffic simulations \cite{thodi2022incorporating}, the simulated dataset may not accurately represent road segments in the real world, depending on the quality of calibrations.
To mitigate the reliance on complete training data, physics-informed deep learning approaches assisted by physical models have conducted successful trials in TSE \cite{huang2020physics, shi2021physics, pshi2021physics, rempe2021estimating, zhao2023observer, zhang2024physics, lu2023physics}. 
However, under conditions of sparse data, the performance of the physics-informed deep learning method may be sensitive to the trade-off between model-driven and data-driven components, making reliable training greatly challenging.


Alternatively, low-rank matrix/tensor completion, a data-efficient grid-based data-driven approach, has emerged to deal with limited data scenarios and achieved promising results in the TSE domain using only sparse observations \cite{shao2018license, tang2020tensor, yu2020urban, wang2023low, nie2022correlating}.
Based on the spatiotemporal grid/matrix, the basic idea of this approach is to recover the spatiotemporal traffic state by representing spatiotemporal traffic dynamic dependencies with algebraic low-rankness. 
For example, \citet{wang2023low} transformed the traffic state matrix into a fourth-order Hankel tensor and applied low-rank matrix completion on the unfolded matrix to recover spatiotemporal traffic speeds using limited vehicle trajectory data. \citet{nie2022correlating} organized spatiotemporal traffic speeds into a tensor and implemented spatiotemporal traffic speeds kriging by graph-embedded tensor completion. However, these pure data-driven low-rank representation methods
may degrade under extremely sparse data environments (e.g. 3$\%$ or less vehicle trajectories).

Focusing on online applications, there are streaming-data-driven methods that only use streaming data (e.g., real-time data) \cite{seo2015estimation, seo2017traffic, han2021estimation, kyriacou2022bayesian}. 
These methods rely less on prior knowledge, thereby demonstrating high robustness to uncertain phenomena and unpredictable incidents.
In addition to conventional fixed and mobile sensor data,  various types of interesting streaming data are also utilized in this category, including extended floating car data (xFCD) that can measure space and time headway \cite{seo2015estimation, kyriacou2022bayesian}, and unmanned aerial vehicle (UAV) data that can provide fast and accurate traffic state observations at any desired locations in multiple travel directions \cite{ke2016real,  theocharides2023towards, ke2018real, theocharides2024real}.
However, a large amount of streaming data is usually required for streaming-data-driven methods to provide accurate state estimations.

\subsection{Research Challenges and Contributions}

Despite the fact that grid-based data-driven methods have achieved high precision in previous literature, researchers continuously contribute to this branch by tackling the following three major challenges: 

\textbf{C1: consistency with backward wave propagation.} 
Previous research has highlighted the advantages of modeling spatiotemporal traffic characteristics along the direction of backward waves, which propagate obliquely \cite{newell1993simplified, laval2011hysteresis}.
However, most Traffic State Estimation (TSE) methods typically use an orthogonal grid-based approach as shown in Fig. \ref{fig: Illustration}(a), 
leading to inconsistencies with the actual propagation of non-orthogonal backward traffic waves. As a result, these inconsistencies cause inhomogeneous traffic states within certain grids, e.g., cells A and B in Fig. \ref{fig: Illustration}(a), potentially introducing biased entries for the TSE and diminishing its accuracy \cite{thodi2022incorporating, wu2023traffic, tsanakas2022generating}. Furthermore, under extremely sparse data environments, constructing the TSM with orthogonal grids may lead to the entire column-missing problem, which may weaken the performance of pure data-driven models depending on column-wise algebra similarity \cite{wang2023low, ma2021high}. Recognizing the limitations of orthogonal grids, \citet{he2019constructing} proposed oblique grids for better alignment with traffic wave propagation, enhancing the segment-level travel time estimation accuracy. For the spatiotemporal grid-level estimation (the focus of this study), they utilized a simple neighborhood-based imputation method, which becomes less effective when significant data is missing.  Additionally, their approach was limited by relatively low estimation resolutions.

\textbf{C2: robustness to corrupted input data.}
The TSE model can be degraded when encountering unfavorable conditions such as noisy or corrupted measurement, emphasizing the robustness requirements against data noise and corruption. The previous works mainly focused on the former and enhanced their model robustness by characterizing the uncertainty caused by stochastic disturbances in TSE \cite{yuan2021macroscopic,wu2023traffic}. However, random data corruption that does not follow Gaussian distribution can also be problematic. Though data pre-processing methods are usually effective in removing these corrupted observations, they might inadvertently filter out genuine observations that are crucial for accurate traffic state estimation, depending on hyper-parameter selection, e.g., filtering threshold. To ensure that all potentially valuable information is utilized for accurate state estimation, a reliable model that is robust to corrupted raw data without destroying its integrity is desirable for TSE.

\textbf{C3: computational complexity.} 
The computational complexity of exiting grid-based data-driven methods is not only related to the problem size (i.e., temporal and spatial length of reconstructed area) but also positively correlated with other variables, such as the number of observations \cite{treiber2011reconstructing, yang2022generalized} and model hyperparameters \cite{wang2023low}, bringing overwhelming computational costs for TSE. 
For large-scale TSE applications with significant problem sizes, it is practically essential to develop an efficient model with no additional scenario-dependent or parameter-induced computational complexity.


The existing studies have attempted to handle one or two of the above challenges. In this study, we propose a tailored matrix completion approach that simultaneously tackles all these three issues.
\textbf{To address the C1}, we integrate traffic wave priors into a customized low-rank matrix completion model based on the oblique grid-modeling approach by He et al. \cite{he2019constructing}.  The differences between their studies and our work are as follows.
First, given oblique grids, instead of exploiting the enhanced traffic state homogeneity only,  we further leverage the enhanced algebraic low-rankness inherent in the traffic state matrix, significantly improving TSE accuracy, especially under severe data scarcity conditions. 
Second,  He et al. \cite{he2019constructing} utilized a simple interpolation-based imputation to estimate traffic states with low resolutions ranging from 150m/90s to 50m/30s, while our study proposes a tailored low-rank approach capable of estimating high-resolution states at 3m/5s, addressing greater challenges with an 88$\%$ rate of empty cells compared to 21$\%$ in the prior work. 
(2)	\textbf{To tackle the C2}, we design an anomaly-tolerance module to accommodate potentially corrupted traffic state observations. Specifically, we assume the ubiquitous data corruptions are randomly and sparsely distributed, and treat the corrupted data detection as a sparse matrix completion problem.
(3)	\textbf{To respond to the C3}, we employ a simple and efficient matrix completion, in which the per-iteration computational complexity is only related to the temporal and spatial length of the TSE reconstructed area.

The contributions of this paper are summarized as follows:

\begin{enumerate}

    \item A traffic wave-inspired low-rank model is tailored for traffic state estimation, in which an oblique grid-based matrix is designed to enhance the low-rank nature within the traffic states and thereby helps to proficiently capture spatiotemporal traffic state dependencies.

    \item	An anomaly-tolerant module is developed to accommodate corrupted data input in robust traffic state estimation, without requiring additional data pre-processing procedures.

    \item Theoretical computational complexity analysis and empirical running time evidence prove the efficiency of the proposed method.
    Numerous experiment results also demonstrate its superior estimation accuracy and robustness.

\end{enumerate}

The remainder of this paper is organized as follows. Section \ref{sec: Preliminaries} gives some basic notations and defines the traffic speed estimation problem. Section \ref{sec: Methodology} formulates the proposed model and derives the associated solving algorithm. Section \ref{sec: Experiments} implements experiments on a real-world traffic dataset and presents the results.  Section \ref{sec: discussions} presents further discussions. Finally, Section \ref{sec: conclusions} concludes this paper and provides future research directions.

\begin{figure*}[t]
  \begin{center}
\includegraphics[width=7.1in]{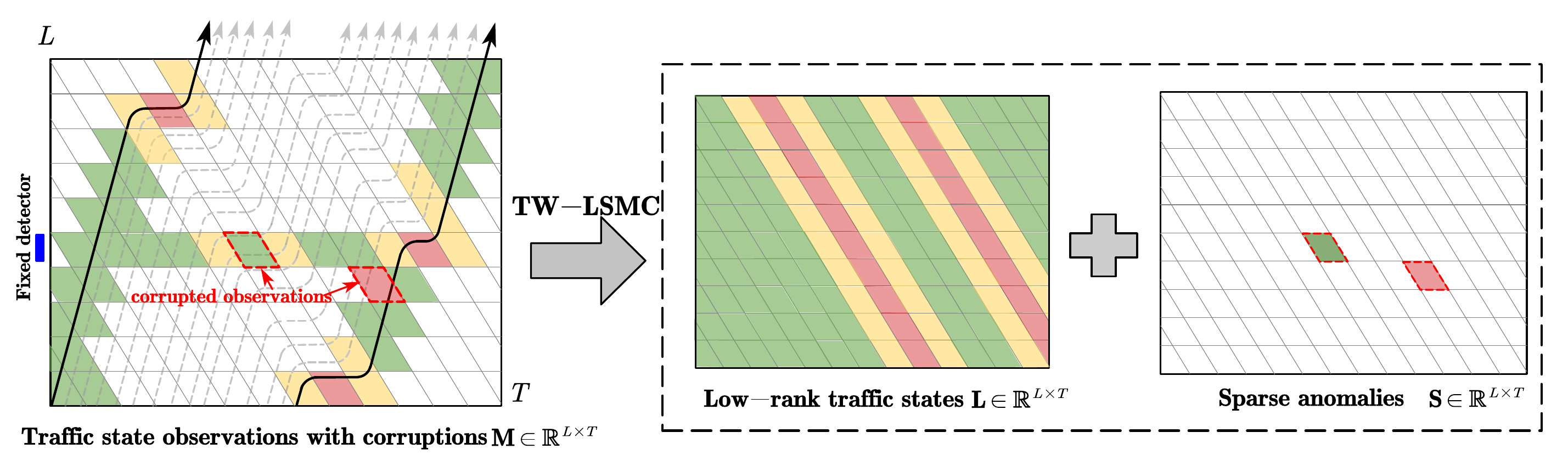}
\caption{Illustration of the proposed method. An oblique grid-based traffic state matrix is constructed (subsection \ref{subsec: matrix construction}) using incomplete and corrupted traffic state observations, and then a low-rank and sparse matrix completion model (subsection \ref{subsec: TSE tailored LRR model}) is applied to recover the complete low-rank spatiotemporal traffic state and to simultaneously detect potential sparse corrupted/anomaly data.}
\label{fig: Illustration-TSLR}
  \end{center}
\end{figure*}

\section{Preliminaries} \label{sec: Preliminaries}

\subsection{Notations}\label{subsec: Notations}
We use lowercase letters to denote scalars, e.g., $a\in \mathbb{R}$, boldface lowercase letters to denote vectors, e.g., $\boldsymbol{a}\in \mathbb{R} ^n$, boldface capital letters to denote matrices, e.g., $\mathbf{A}\in \mathbb{R} ^{n_1\times n_2}$, and Euler script letters to denote third-order tensors, e.g., $\mathcal{A} \in \mathbb{R} ^{n_1\times n_2\times n_3}$. 
Given a matrix $\mathbf{X}\in \mathbb{R} ^{n_1\times n_2}$, the matrix nuclear norm is denoted as $\left\| \mathbf{X} \right\| _*=\sum\nolimits_{i=1}^{min\left( n_1,n_2 \right)}{\sigma _i\left( \mathbf{X} \right)}$, where $\sigma _i\left( \mathbf{X} \right) $ is the $i$th largest singular value of $\mathbf{X}$, and the Frobenius norm is defined as $\left\| \mathbf{X} \right\| _F=\sqrt{\sum\nolimits_{i=1}^{n_1}{\sum\nolimits_{j=1}^{n_2}{x_{ij}^{2}}}}$. The inner product between two matrices of the same size is $\left< \mathbf{A}, \mathbf{B} \right> =\mathrm{Tr}\left( \mathbf{A}^{\mathsf{T}}\mathbf{B} \right) =\sqrt{\sum\nolimits_{i=1}^{n_1}{\sum\nolimits_{j=1}^{n_2}{a_{ij}b_{ij}}}}$, where $\mathrm{Tr}\left( \cdot \right) $ is the matrix trace.

\subsection{Problem description}

We aim to estimate freeway traffic speeds at fixed 5-second intervals over extended periods, using trajectory data collected from mobile sensors such as connected vehicles (CVs).  For a single lane of the freeway segment, traffic speed variables are collected in the spatiotemporal domain $S\times W$, where $S$ is segment length and $W$ is time window length. Given predefined spatial resolution $\varDelta s$  and temporal resolution $\varDelta t$, we can transform the traffic state measurements into a discrete space with matrix representation $\mathbf{M}\in \mathbb{R} ^{L\times T}$, where  $L=S/\varDelta s$ and $T=W/\varDelta t$. The value of each cell is the average traffic state variable of that cell (detailed descriptions are introduced in subsection \ref{subsec: matrix construction}).

The observed traffic state matrix $\mathbf{M}$  is usually incomplete and highly sparse since the data from fixed and mobile sensors have limited spatiotemporal coverage.  In addition, the observed entries in $\mathbf{M}$  may also contain corrupted data due to false records and communication failures, which further complicates the requirements of model robustness.
To this end, we here differentiate such two levels of TSE requirements by defining two specific tasks as follows 

\begin{itemize}
  \item Traffic state estimation (TSE): to reconstruct the precise and complete spatiotemporal traffic state from \textbf{sparse but pure} observations.

  \item Robust traffic state estimation (RTSE): to simultaneously identify the potentially corrupted data and recover precise and complete spatiotemporal traffic state from \textbf{sparse and corrupted  (also called anomaly \cite{wang2021diagnosing})} observations.
\end{itemize}
Note that the term "traffic state" is used to refer to the speed states specifically in this study.

\section{Methodology} \label{sec: Methodology}

In this section, we propose an efficient and robust approach for freeway traffic state estimation. Firstly, \textbf{regarding C1},  we incorporate backward wave priors to construct an oblique grid-based traffic state matrix in subsection \ref{subsec: matrix construction}. After that, \textbf{regarding C2},  we build a robust matrix completion (MC) model to recover accurate traffic state from sparse and anomaly-corrupted data in subsection \ref{subsec: TSE tailored LRR model}. Then, an Alternating Direction Method of Multipliers (ADMM)-based iterative solving framework is elaborated in subsection \ref{subsec: solving framework}. Finally, \textbf{regarding C3}, we analyze the computational complexity of the proposed model in subsection \ref{subsec: Computational complexity}.

\subsection{Oblique grid-based traffic state matrix construction (C1)}
\label{subsec: matrix construction}

\begin{figure}[h]
  \begin{center}
\includegraphics[width=2.5in]{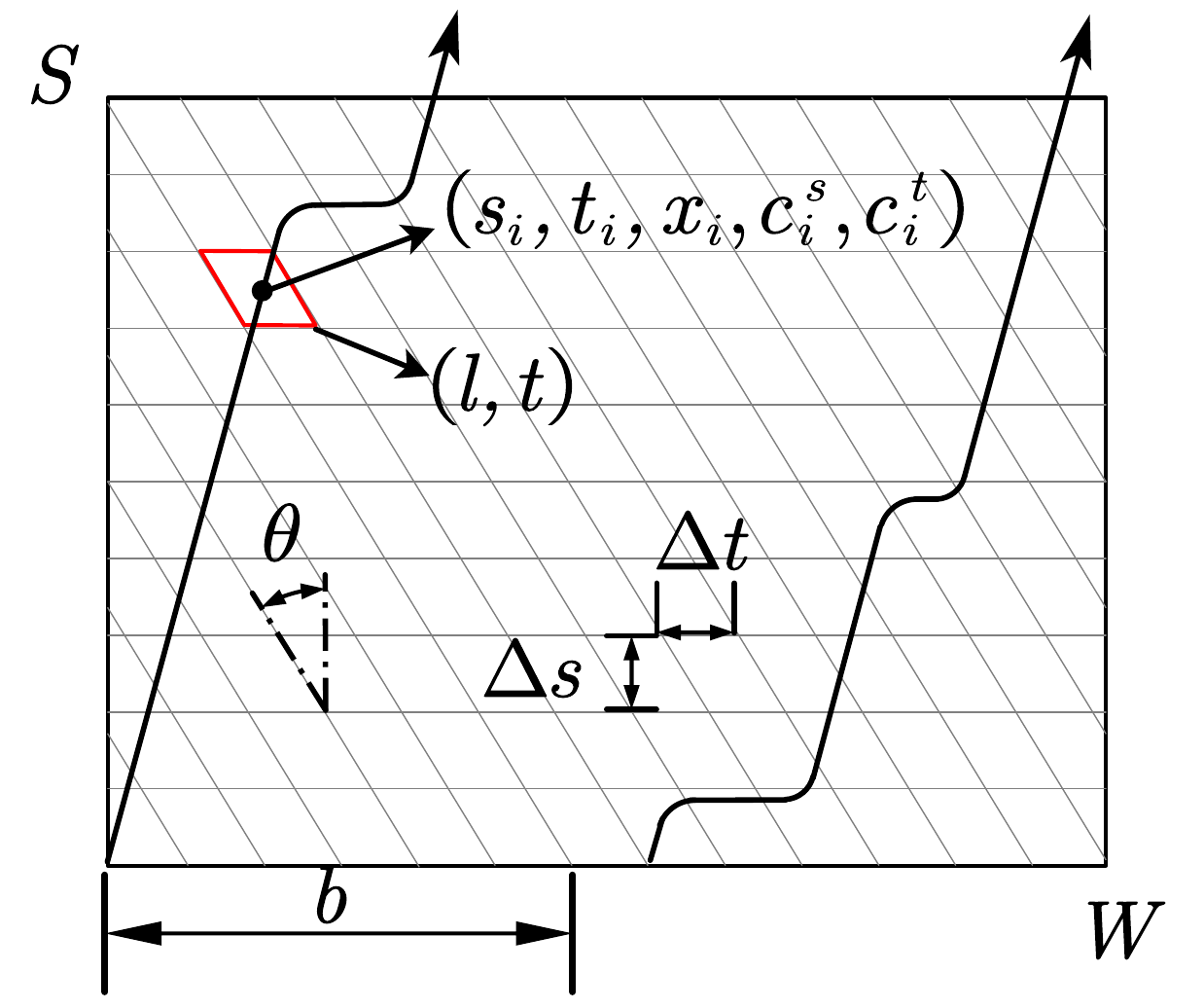}
\caption{Illustration of constructing an oblique grid-based traffic state matrix.}
\label{fig: Illustration-matrix construction}
  \end{center}
\end{figure}

To construct the spatiotemporal traffic state matrix (TSM), an intuitive idea is to virtually partition a spatiotemporal plane into orthogonal grids (see Fig.\ref{fig: Illustration} (a)), introducing the inconsistency mentioned in the C1. 
To alleviate these inconsistencies, \citet{he2019constructing} proposed using non-rectangular/oblique grids to construct spatiotemporal diagrams and proved its advantages over using conventional rectangular grids by the improved results of segment-level travel time estimation accuracy.
However, for the fine-grained cell-level traffic state estimation (the focus of this study), they adopted a simple neighborhood-based iterative imputation method to fill empty cells in the spatiotemporal diagram, which may be sharply degraded when a large portion of cells are missing.
\textbf{To address the C1 in TSE}, based on the prior work, we follow the idea of oblique grids and extend it to fine-grained (e.g., 3m/5s) TSE under extreme missing conditions by constructing an oblique grid-based traffic state matrix, where the inclines of the left and right edges are aligned with the backward wave speed, as shown in Fig. \ref{fig: Illustration-matrix construction}.

Given traffic state observations  $\left( s_i,t_i,x_i \right), i=1,..,N $, where the $s_i$  and $t_i$  are the spatial and temporal coordinates of traffic state variable $x_i$, we aim to construct a TSM along the direction of backward traffic wave to ensure the homogeneity within each entry of the TSM. The first step is to determine the spatial and temporal cell index  $c_i^s$ and  $c_i^t$  that each observation belongs to
\begin{align}
  &c_{i}^{s}=s_i|\varDelta s,
  \\
  &c_{i}^{t}=\left( t_i-\left( b-s_i\cdot \tan \left( \theta \right) \right) \right) |\varDelta t,
\end{align}
where $\varDelta s$ and $\varDelta t$ are the spatial resolution and temporal resolution used in TSM construction,  $\theta$ is the inclined angle of the backward wave, and $\theta =arccot\left( v/3.6 \right) $, where $v$ is the backward wave speed that generally ranges from -10 km/h to -20 km/h \cite{chen2014periodicity, he2019constructing},  $b$ is the intercept constant, and $b = S\cdot \tan \left( \theta \right) $, where $S$ is the spatial length of the target segment. The representative traffic state values of each cell $\left( l,t \right) $ are calculated by averaging the observed traffic state values within the cell 
\begin{align}
  \bar{x}_{l,t}=\frac{1}{N_{l,t}}\sum_{c_{i}^{s}=l,c_{i}^{t}=t}{x_i},
\end{align}
where $N_{l,t}$ is the total number of observation points within the cell $\left( l,t \right) $.

\subsection{Low-rank and Sparse Matrix Completion (C2)}
\label{subsec: TSE tailored LRR model}

Traffic states exhibit distinct spatiotemporal dependencies, such as temporal periodicity and spatial propagation characteristics shown in Fig. \ref{fig: Illustration}. 
By constructing the traffic state matrix (TSM) with oblique grids illustrated in subsection \ref{subsec: matrix construction}, the highly correlated traffic states along the backward wave direction are strategically aligned into the same matrix column. This alignment adeptly transforms the traffic state correlations such as temporal recurrences and spatial dependencies into the algebraic low-rankness of a matrix, i.e., the column-wise or row-wise similarity.  
In other words, the low-rankness of the TSM is enhanced using oblique-grid-based modeling,  ensuring a low-rank representation method to proficiently capture the spatiotemporal correlations inherent in traffic states. This approach enables the precise reconstruction of traffic states from sparse observations by reformulating the TSE problem as a low-rank matrix completion task.

Specifically, given a partially observed traffic state matrix, the low-rank matrix completion aims to estimate the target  complete state matrix $\mathbf{L}$ by minimizing its algebraic rank
\begin{align}
  \underset{\mathbf{L}}\min\,\,\mathrm{rank}\left( \mathbf{L} \right) \,\,s.t.\,P_{\Omega}\left( \mathbf{L} \right) =P_{\Omega}\left( \mathbf{M} \right) ,
\label{eq: rank mini}
\end{align}
where $\mathbf{M}$ is the partially observed traffic state matrix, and the constraint ensures that the values of $\mathbf{L}$ and $\mathbf{M}$ are consistent at the observation set $\Omega$. Considering the rank minimization in Eq. \eqref{eq: rank mini} is an NP-hard problem, several convex and non-convex surrogate functions are applied to ensure computational feasibility. In this study, we employ nonconvex truncated nuclear norm \cite{hu2012fast} as the rank function, the problem in Eq. \eqref{eq: rank mini} can be rewritten as
\begin{align}
  \underset{\mathbf{L}}\min\,\,\left\| \mathbf{L} \right\| _{r,*}\,\,\,\, s.t.\,\,P_{\Omega}\left( \mathbf{L} \right) =P_{\Omega}\left( \mathbf{M} \right) ,
\label{eq: nonconvex model}
\end{align}
where $\left\| \mathbf{X} \right\| _{r, *}$ is the truncated nuclear norm of matrix $\mathbf{X}$.

However, as aforementioned in C2, the potential data corruption in traffic state observations may adversely affect the model performance. \textbf{To address the C2}, we assume the data corruptions are randomly and sparsely distributed and introduce a sparse matrix $\mathbf{S}$ to accommodate these corruptions. 
A robust \textbf{T}raffic \textbf{W}ave based \textbf{L}ow-rank and \textbf{S}parse \textbf{M}atrix \textbf{C}ompletion model (TW-LSMC) is presented as 
\begin{align}
    \underset{\mathbf{L},\mathbf{S}}\min\,\,\left\| \mathbf{L} \right\| _{r,*}+\lambda \left\| \mathbf{S} \right\| _1\,\,s.t.\,P_{\Omega}\left( \mathbf{L}+\mathbf{S} \right) =P_{\Omega}\left( \mathbf{M} \right),
\label{eq: TW-LSMC}
\end{align}
where $\left\| \mathbf{L} \right\| _{r,*}$ is the truncated nuclear norm of low-rank matrix $\mathbf{L}$, and $\left\| \mathbf{S} \right\| _1$ is the $l_1$ norm of sparse matrix $\mathbf{S}$,  $\lambda$ is a weight parameter that balances the trade-off between low-rank and sparse regularization. In the proposed model, the traffic state observations are represented as a combination of low-rank structural and sparse anomaly components to simultaneously recover the complete and accurate traffic state and detect the anomaly data.

\subsection{Iterative solving framework using ADMM (C2)}
\label{subsec: solving framework}
To reserve the original observed information in each iteration, we do not directly update the observation matrix $\mathbf{M}$ but introduce an auxiliary variable $\mathbf{W}$ to conduct the update and transfer the observations from $\mathbf{M}$ to $\mathbf{L}$ and $\mathbf{S}$. The model in Eq.\eqref{eq: TW-LSMC} is reformulated as 
\begin{align}
  &\underset{\mathbf{L},\mathbf{S}}\min~\left\| \mathbf{L} \right\| _{r,*}+\lambda \left\| \mathbf{S} \right\| _1,
  \nonumber\\
  &s.t.~\mathbf{W}=\mathbf{L}+\mathbf{S}, P_{\Omega}\left( \mathbf{W} \right) =P_{\Omega}\left( \mathbf{M} \right). 
\label{eq: extend TW-LSMC}
\end{align}
To cope with the equal constraint, the augmented Lagrangian function of our TW-LSMC model is written as
\begin{align}\label{eq: lagrange function}
&\mathcal{L} \left( \mathbf{L},\mathbf{S},\mathbf{W},\mathbf{Y} \right) =\left\| \mathbf{L} \right\| _{r,*}+\left\| \mathbf{S} \right\| _{1}+\frac{\rho}{2}\left\| \mathbf{W}-\mathbf{L}-\mathbf{S} \right\| _{F}^{2}
\\ \nonumber
&+\left< \mathbf{Y},\mathbf{W}-\mathbf{L}-\mathbf{S} \right>, ~s.t.\,\,P_{\Omega}\left( \mathbf{W} \right) =P_{\Omega}\left( \mathbf{M} \right),
\end{align}
where $\left< \cdot ,\cdot \right>$ indicates the inner product, $\mathbf{Y}\in \mathbb{R} ^{n_1\times n_2}$ denotes the Lagrangian multiplier and $\rho>0$ represents the penalty parameter. According to the ADMM framework, the minimization of our model can be decomposed into iteratively solving the following three subproblems:
\begin{align}
    \mathbf{L}^{l+1}&=\underset{\mathbf{L}}{\argmin}\,\,\mathcal{L} \left( \mathbf{L},\mathbf{S}^l,\mathbf{W}^l, \mathbf{Y}^l \right), \label{eq: update L}
    \\
    \mathbf{S}^{l+1}&=\underset{\mathbf{S}}{\argmin}\,\,\mathcal{L} \left( \mathbf{L}^{l+1},\mathbf{S},\mathbf{W}^l,\mathbf{Y}^l \right), \label{eq: update S}
    \\
    \mathbf{W}^{l+1}&=\underset{\mathbf{W}}{\argmin}\,\,\mathcal{L} \left( \mathbf{L}^{l+1},\mathbf{S}^{l+1},\mathbf{W},\mathbf{Y}^l \right), \label{eq: update W}
    \\
    \mathbf{Y}^{l+1}&=\mathbf{Y}^l+\rho \left( \mathbf{W}^{l+1}-\mathbf{L}^{l+1}-\mathbf{S}^{l+1} \right), \label{eq: update Y}
\end{align}
where $l$ denotes the $l$-th iteration, and the three variables $\mathbf{L}, \mathbf{S}, \mathbf{W} $ are alternatively updated in each iteration until convergence. The detailed solutions of Eq. \eqref{eq: update L}, Eq. \eqref{eq: update S}, and Eq. \eqref{eq: update W} are given in the following subsections.
The pseudocode of TW-LSMC numerical solution is summarized in Algorithm \ref{ag: TSLR}.

\subsubsection{Update Variable $\mathbf{L}$} Removing the irrelevant terms, the $\mathbf{L}$ subproblem is written as 
\begin{align}
    \mathbf{L}^{l+1}&=\underset{\mathbf{L}}{\argmin}\left\| \mathbf{X} \right\| _{r,*}+\frac{\rho}{2}\left\| \mathbf{W}^l-\mathbf{L}-\mathbf{S}^l \right\| _{F}^{2}-\left< \mathbf{Y}^l,\mathbf{L} \right> 
    \nonumber\\
    &=\underset{\mathbf{L}}{\argmin}\left\| \mathbf{L} \right\| _{r,*}+\frac{\rho}{2}\left\| \mathbf{L}-\left( \mathbf{W}^l-\mathbf{S}^l+\frac{\mathbf{Y}^l}{\rho} \right) \right\| _{F}^{2}
    \nonumber\\
    &=\mathcal{D} _r\left( \mathbf{W}^l-\mathbf{S}^l+\frac{\mathbf{Y}^l}{\rho} \right), 
\label{eq: up_date L}
\end{align}
where $\mathcal{D} _r$ is the weighted singular value thresholding operator as shown in \textbf{Lemma 1}.

\vspace{0.5em}
\noindent\textbf{Lemma 1.} \citep{hu2012fast} \textit{For any $\rho >0$, $\mat{Z}\in \mathbb{R} ^{m\times n} $, and $r\in \mathbb{N} _+$ where $r<\min \left\{ m,n \right\} $, an optimal solution to the truncated nuclear norm minimization problem
\begin{align}
    \underset{\mathbf{X}}\min  \left\| \mathbf{X} \right\| _{r,*}+\frac{\rho}{2}\left\| \mathbf{X}-\mathbf{Z} \right\| _{F}^{2},
\end{align} 
is given by the weighted singular value thresholding 
\begin{align}
    \mathscr{D} _{r,1/\rho}\left( \mathbf{Z} \right) =\mathbf{U}\mathrm{diag}\left( \left[ \boldsymbol{\sigma }-\mathbbm{1}\cdot 1 /\rho \right] _+ \right) \mathbf{V}^{\mathsf{T}},
\end{align}
where $\mathbf{U}\mathrm{diag}\left( \boldsymbol{\sigma } \right) \mathbf{V}^{\mathsf{T}}$ is the singular value decomposition of $\mat{Z}$, $\left[ \cdot \right] _+$ denotes the positive truncation at $0$ which  satisfies $\left[ \sigma -1 /\rho \right] _+=\max \left\{ \sigma -1 /\rho ,0 \right\} $, $\mathbbm{1}\in \left\{ 0,1 \right\} ^{\min \left\{ m,n \right\}}$ is a binary indicator vector whose first $r$ entries are $0$ and other entries are $1$.}

\subsubsection{Update Variable $\mathbf{S}$} Specifically, the $\mathbf{S}$ subproblem is written as
\begin{align}
    \mathbf{S}^{l+1}&=\underset{\mathbf{S}}{\argmin}~\lambda \left\| \mathbf{S} \right\| _1+\frac{\rho}{2}\left\| \mathbf{W}^l-\mathbf{L}^{l+1}-\mathbf{S} \right\| _{F}^{2}-\left< \mathbf{Y}^l,\mathbf{S} \right> 
    \nonumber\\
    &=\underset{\mathbf{S}}{\argmin}~\lambda \left\| \mathbf{S} \right\| _{1}+\frac{\rho}{2}\left\| \mathbf{S}-\mathbf{H} \right\| _{F}^{2}
    \nonumber\\
    &=\mathrm{sgn} \left( \mathbf{H} \right) \circ \max \left\{ \left| \mathbf{H} \right|-\frac{\lambda}{\rho},\,\,0 \right\} ,
\label{eq: up_date S}
\end{align}
where $\mathbf{H}=\mathbf{W}^l-\mathbf{L}^{l+1}+\frac{\mathbf{Y}^l}{\rho}$,  $\circ$ indicates the point-wise product, and the $\mathrm{sgn} \left( \cdot \right) $ denotes the signum function, i.e.,
\begin{align}
\mathrm{sgn}\left( x \right) =\begin{cases}
	1  ~~  &\mathrm{if} ~ x>0,\\
	0  ~~ &\mathrm{if} ~ x=0,\\
	-1 ~~ &\mathrm{if} ~ x<0.\\
\end{cases}~~~~~~~~~~~~~~~~~~~~
\label{eq:problem3.2.10}
\end{align}

\subsubsection{Update Variable $\mathbf{W}$}
The $\mathbf{W}$ sub-problem is a set of unconstrained quadratic equations element-wise. Therefore, the closed-form solution is obtained as
\begin{align}
    \mathbf{W}^{l+1}&=\underset{\mathbf{W}}{\argmin}~\frac{\rho}{2}\left\| \mathbf{W}-\mathbf{L}^{l+1}-\mathbf{S}^{l+1} \right\| _{F}^{2}+\left< \mathbf{Y}^{l},\mathbf{W} \right> \,\,
    \nonumber\\
    &=\underset{\mathbf{W}}{\argmin}~\frac{\rho}{2}\left\| \mathbf{W}-\left( \mathbf{L}^{l+1}+\mathbf{S}^{l+1}-\frac{\mathbf{Y}^{l}}{\rho} \right) \right\| _{F}^{2}\,\,
    \nonumber\\
    &=\mathbf{L}^{l+1}+\mathbf{S}^{l+1}-\frac{\mathbf{Y}^{l}}{\rho},
\label{eq: up_date W}
\end{align}
and the following transformation holds: 
\begin{align}
P_{\Omega}\left( \mathbf{W}^{l+1} \right) =P_{\Omega}\left( \mathbf{M} \right),
\label{eq: W constraint}
\end{align}
where ${\Omega}$ is the observation set of spatiotemporal traffic state.

\begin{algorithm}
  \caption{Numerical solution of Eq. \eqref{eq: lagrange function} via ADMM}
  \KwIn{The partially measured traffic state matrix $\mathbf{M}$, weight parameter $\lambda$, truncated parameter $r$.}
  \KwOut{The recovered low-rank traffic state matrix $\mathbf{L}$, and sparse anomaly matrix $\mathbf{S}$}
  \textbf{Initialization:} $\rho =10^{-4},\varepsilon =10^{-4},l=1,
  \mathbf{L}=\mathbf{W}=\mathbf{M},\mathbf{M}_{\Omega ^-}=\mathrm{mean}\left( \mathbf{M}_{\Omega} \right), \mathbf{S}=\mathbf{O}^{n_1\times n_2}$, where $\mathbf{O}$ denotes a matrix with all entries equal to zero  \;
  \LinesNumbered
  \SetAlgoVlined
  \While{not converged}{
  Update $\mathbf{L}^{l+1}$ via Eq. \eqref{eq: up_date L} \;
  Update $\mathbf{S}^{l+1}$ via Eq. \eqref{eq: up_date S} \;
  Update $\mathbf{W}^{l+1}$ via Eq. \eqref{eq: up_date W} and \eqref{eq: W constraint}\;
  Update $\mathbf{Y}^{l+1}$ via Eq. \eqref{eq: update Y}\;
  Calculate $\frac{\left\| \mathbf{L}^{l+1}-\mathbf{L}^l \right\| _{\mathrm{F}}}{\left\| \mathbf{L}_{\Omega}^{0} \right\| _{\mathrm{F}}}<\epsilon $\;
  $l=l+1$
  }
  \label{ag: TSLR}
\end{algorithm}

\subsection{Computational complexity (C3)}
\label{subsec: Computational complexity}
The computational complexity of the Algorithm \ref{ag: TSLR} is dominated by the update of low-rank matrix $\mathbf{L}\in \mathbb{R} ^{L\times T}$, which involves a matrix truncated nuclear norm minimization problem with respect to matrix $\mathbf{L}$.
Specifically, the $\mathbf{L}$ subproblem only needs to solve a singular value decomposition (SVD) of $L\times T$ matrix in each iteration, contributing to a per-iteration computation complexity of $\mathcal{O} \left( L^2T \right) $ when $L<T$ .  By denoting the number of iterations by $k$, we can obtain that the computational complexity of Algorithm \ref{ag: TSLR} is $ \mathcal{O} \left(k L^2T \right) $.

\section{Experiments}  \label{sec: Experiments}

In this section, we evaluate our proposed TW-LSMC method on real-world traffic dataset in comparison with state-of-the-art methods, which are summarized to answer the following research questions (RQs): 
\begin{itemize}
  \item RQ1 (\ref{subsec: performance evaluation: TSE}): How about the performance of the proposed TW-LSMC in sparse data environments? 
  \item RQ2 (\ref{subsec: performance evaluation: RTSE}): How about the performance of the proposed TW-LSMC with corrupted data input?
  \item RQ3 (\ref{subsec: Sensitivity analysis}): How does the wave speed parameter of TW-LSMC affect the TSE performance?
  \item RQ4 (\ref{subsec: Ablation study}): How do different model components contribute to model performance?
  \item RQ5 (\ref{subsec: Computation Performance}): How about the computational efficiency of the proposed model compared to existing SOTA methods?
\end{itemize}

\subsection{Data description and corrupted data generation}

\begin{figure*}[t]
  \begin{center}
  \includegraphics[width=7in]{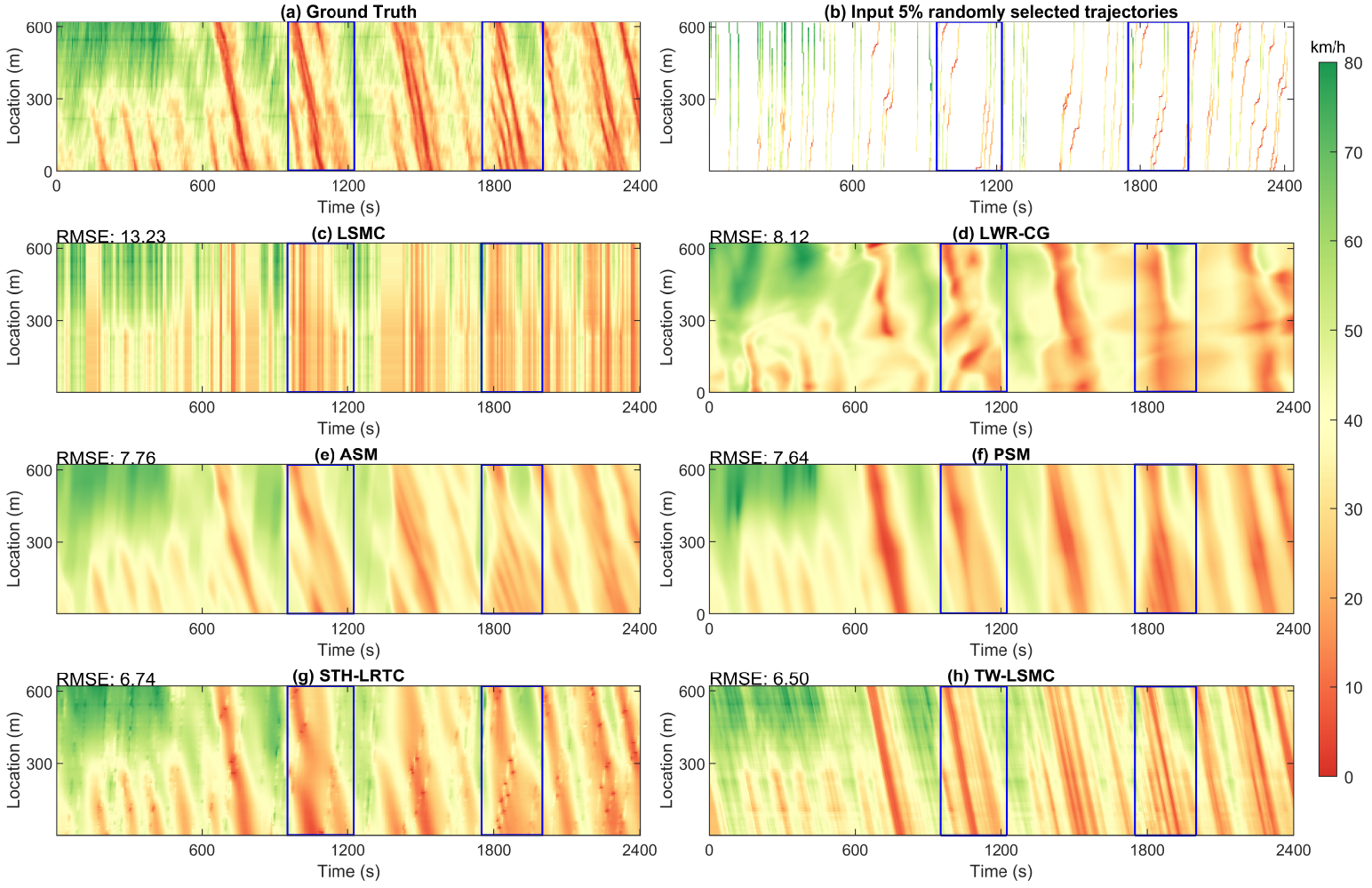}
  \captionsetup{font={small}}
  \caption{A TSE experiment on the NGSIM dataset: (a) The ground truth traffic speed; (b) The observed traffic speed from 5$\%$ randomly selected vehicle trajectories;  (c) The estimation result by LSMC;  (d) The estimation result by LWR-CG; (e) The estimation result by ASM; (f) The estimation result by PSM; (g) The estimation results by STH-LRTC; (h) The estimation results by the proposed TW-LSMC.}
  \label{fig: Visualization-TSE}
  \end{center}
\end{figure*}

In this study, we use vehicle trajectories extracted from video cameras on lane 2 of US Highway 101 of the NGSIM dataset.
Similar to the previous work by \citet{wang2023low},  our experiments cover a segment of 621 meters, and the test duration is 2400 seconds. We focus on the traffic state with a resolution of 3 meters and 5 seconds, where the traffic state is defined as the average vehicle speed in each grid cell. Consequently, the spatiotemporal size of the traffic state matrix is  $207 \times 480$. The traffic speed maps of the entire dataset are shown in Fig. \ref{fig: Visualization-TSE} (a).

To evaluate the model performance on robust traffic state estimation, we design two types of non-Gaussian data corruption that may adversely affect the TSE performance:
\begin{itemize}
  \item Type I: the observed data under the free-flow state are tampered to the jam waves/stop-and-go waves state. 
  \item Type II: the observed data under the jam waves/stop-and-go waves state are tampered to the free-flow state. 
\end{itemize}
These two types of corruption introduce false information and can greatly affect the estimation of the surrounding traffic state. We define the tampered speed of two types of corruption as follows
\begin{align}
  v_{\mathrm{I}}=v_f-50, \label{eq: type I corruption}
  \\
  v_{\mathrm{II}}=v_{\mathrm{c}}+80, \label{eq: type II corruption}
\end{align}where the $v_f\geqslant 50$ km/h  and  $v_c\leqslant 5$ km/h are the actual speed observations under free-flow and jam waves/stop-and-go waves state \cite{karl2019automated}.

\subsection{Baseline models and evaluation metrics}
We compared the proposed TW-LSMC model with the following six alternative methods:

\begin{itemize}
  \item \textbf{LSMC} (Low-rank and Sparse Matrix Completion, \cite{candes2011robust}): A rectangular grid-based low-rank and sparse matrix completion method with truncated nuclear norm minimization \cite{hu2012fast} and $l_1$ norm minimization.
  \item \textbf{LWR-CG} (LWR model-based Computational Graph, \cite{lu2023physics}): A multi-source data compatible computational graph approach incorporating the LWR model \cite{lighthill1955kinematic, richards1956shock}, three-detector model \cite{newell1993simplified}, and fluid queue model for traffic state and queue profile joint estimation.  As only vehicle trajectory data is used in this study, the first two physical models are mainly operational. 
  \item \textbf{ASM} (Adaptive Smoothing Method, \cite{treiber2011reconstructing}): a spatiotemporal kernel-weighted method that considers free-flow and congested traffic wave propagation characteristics.
  \item \textbf{SD-EGTF/SD-ASM } (Shear/Oblique Grid-based Discrete Extended Generalised Treiber–Helbing Filter (EGTF), \cite{tsanakas2022generating}): An oblique grid-based EGTF \cite{van2010robust} speed state estimator for virtual vehicle trajectory generation. As only one data source (e.g., vehicle trajectories) is used in this study, the EGTF degrades to the Generalised Treiber–Helbing Filter (i.e., Adaptive Smoothing Method) \cite{treiber2002reconstructing, treiber2011reconstructing}. For clarity, we denote the SD-EGTF as SD-ASM in the following sections.
  \item \textbf{PSM} (Phase-based Smoothing Method, \cite{rempe2017phase}): A kernel-weighted smoothing method based on Kerner's three-phase theory \cite{kerner2009introduction}.
  \item \textbf{STH-LRTC} (Spatiotemporal Hankel Low-Rank Tensor Completion, \cite{wang2023low}): A low-rank tensor completion with the spatiotemporal Hankelization to reconstruct the spatiotemporal traffic speed.

\end{itemize}


The hyperparameters in each model greatly affect the TSE performance. For a fair comparison, the baseline models are fine-tuned. 
For the ASM model, we set the parameters according to the suggested values in \cite{treiber2011reconstructing, wang2023low}. Specifically, the wave speeds are set as $v_f$ = 60 km/h and $v_c$ = -10 km/h, the kernel parameters are $\sigma = 200m, \tau = 10s$, and the weighted parameters are $\Delta V = 10$ km/h and $V_{thr} = 20$ km/h.
For STH-LRTC, the parameter setting $\tau_s=40, \tau_t = 30$ is used to obtain the Hankel tensor in \citet{wang2023low}. However, we find that this setting provides poor estimation results in some cases. According to the parameter grid search results, we set the embedding length that achieved the best overall performance for each scenario in our experiments, as noted in Tab. \ref{tab: TSE performance}. 
For the PSM, we set the speed thresholds $V_{J}^{thr} = 25$km/h, $V_{S}^{thr}=65$ km/h, $V_{F}^{thr}=55$km/h, smoothing directions $V_{J,S}^{dir} = -18$km/h, $V_F^{dir}=70$ km/h, kernel parameters $\tau_{S,F} = 20$s, $\tau _{F,S}^{H}=20$s, $\sigma_{F,S}=100$m as suggested in \citet{rempe2017phase}. For the LWR-CG, we set the weight of partial differential equations (PDE) as 1, the learning parameter as $10^{-4}$, and the number of epochs as 10000. The distributed computing framework is utilized in the separated periods [0s,1200s] and [1200s, 2400s] due to computing memory constraints.
For the proposed method, we use truncated percentage parameter $\theta=0.3$, weighted parameter $\lambda=0.04$, and learning rate control parameter $\rho = 10^{-4}$,  as illustrated in Algorithm \ref{ag: TSLR}.

To guarantee fair comparisons, all experiments are conducted on a desktop with a 3.7 GHz Intel Core i5-9600 K processor and 32 GB of RAM. The STH-LRTC, ASM, and SD-ASM are implemented using Matlab R2018b. The LWR-CG is implemented using Python with TensorFlow-2.10.0. The PSM is coded using Python 3.8 with Numpy-1.19.2 and Pytorch-1.9.0. The LSMC and proposed TW-LSMC are coded in Python 3.8 using NumPy-1.19.2 only. The code is available at https://github.com/heyang49/TW-LSMC.

The partially observed speed data from trajectories are used to recover the full traffic speed in the following TSE and RTSE experiments. Specifically, we randomly select trajectories as input data. We use Root Mean Squared Error (RMSE) and Mean Absolute Error (MAE) as evaluation metrics to evaluate the performance of different models under TSE and RTSE scenarios.
\begin{align}
  &\mathrm{RMSE}=\sqrt{\frac{1}{n}\sum\nolimits_{i=1}^n{\begin{array}{c}
    \left( y_i-\hat{y}_i \right) ^2\\
  \end{array}}},
  \\
  &\mathrm{MAE}=\frac{1}{n}\sum\nolimits_{i=1}^n{\left| y_i-\hat{y}_i \right|},
\end{align}
where $n$  is the number of test data, $y_i$  is the ground truth and  $\hat{y}_i$ is the estimation. Note that the ground truth speed is calculated from all the trajectory points within the grid cell.

\subsection{Traffic State Estimation (RQ1)}
\label{subsec: performance evaluation: TSE}

To assess the TSE model performance, we begin by visualizing the estimation results of the proposed and alternative methods under a 5 \% CV penetration rate scenario using the NGSIM data. Fig. \ref{fig: Visualization-TSE}(a) displays the ground truth traffic speed matrix, depicting intricate traffic dynamics evolution with multiple shockwaves, thereby making it desirable for performance evaluation. Fig. \ref{fig: Visualization-TSE}(b) shows a training dataset chosen from 20 independent experiments, highlighting significant data missing during certain intervals, such as between 950s to 1225s and 1750s to 2000s, which complicates the task for models to accurately reconstruct traffic speeds.

Fig. \ref{fig: Visualization-TSE}(c) visualizes the results using the vanilla LSMC, where LSMC’s state estimates are significantly deficient in the columns that speed observations are entirely missing, primarily because the standard low-rank technique relies heavily on column/row-wise similarities, i.e., algebraic low-rankness. Leveraging partial differentiation equations, the physics-informed LWR-CG method (seen in Fig. \ref{fig: Visualization-TSE}(d)) offers continuous state estimations and depicts congestion patterns, but struggles to precisely reconstitute shockwaves in predominantly missing areas. By applying isotropic smoothing kernels based on the two-phase \cite{newell1993simplified} and three-phase \cite{kerner2009introduction} wave theory, the ASM (Fig. \ref{fig: Visualization-TSE}(e)) and the PSM (Fig. \ref{fig: Visualization-TSE}(f)) reconstruct clearer shockwaves than the LWR-CG. The speed estimations of ASM in the jam area tend to be lower than actual due to the smoothing effects, a limitation mitigated by PSM which offers refined speed estimates. Comparatively, PSM notably outperforms ASM, particularly in the jam and transition areas, owing to its integration of a synchronized flow phase. The STH-LRTC (Fig. \ref{fig: Visualization-TSE}(g)) surpasses both ASM and PSM in accuracy. However, during the period with limited observations (see blue rectangles), both STH-LRTC and smoothing models inadequately estimate shockwaves. In contrast, the proposed TW-LSMC approach showcased in Fig. \ref{fig: Visualization-TSE}(h) successfully reconstructs both major and minor shockwaves with fine-grained features, such as accurate wave lengths, and clear wave boundaries, demonstrating remarkable robustness to sparse data. Driven by an understanding of traffic wave behaviors, the TW-LSMC identifies highly correlated traffic states generated by the same backward wave and builds connections among these states, particularly in distant positions, through a low-rank framework. 

By comparing Fig. \ref{fig: Visualization-TSE}(c), (g) with Fig. \ref{fig: Visualization-TSE}(h), it is evident that the proposed oblique grid-based TW-LSMC adeptly captures distinct traffic propagation characteristics such as stop-and-go shockwaves, which conventional low-rank-based LSMC and STH-LRTC methods cannot model. This leads to remarkable enhancements in estimation accuracy, exemplified by a reduction of 6.73 in RMSE when compared to the rectangular grid-based LSMC. These results confirm the necessity of incorporating traffic wave priors and the effectiveness of the oblique grid in enhancing traffic state low-rankness. Furthermore, the improved low-rankness rendered by the TW-LSMC not only enhances its accuracy over the purely data-driven STH-LRTC approach but also improves its efficiency. Detailed discussions on the theoretical complexity analysis and supporting empirical evidence are provided in subsection \ref{subsec: Computation Performance}. By comparing Fig. \ref{fig: Visualization-TSE}(e), (f) with Fig. \ref{fig: Visualization-TSE}(h), we can observe that incorporating traffic wave priors into two distinct modeling approaches, the low-rank-based TW-LSMC outperforms smoothing-based the ASM and PSM approaches, indicating the superiority of low-rank representation in learning inherent traffic state dependencies.

\begin{figure}[t]
  \begin{center}
  \includegraphics[width=3.3in]{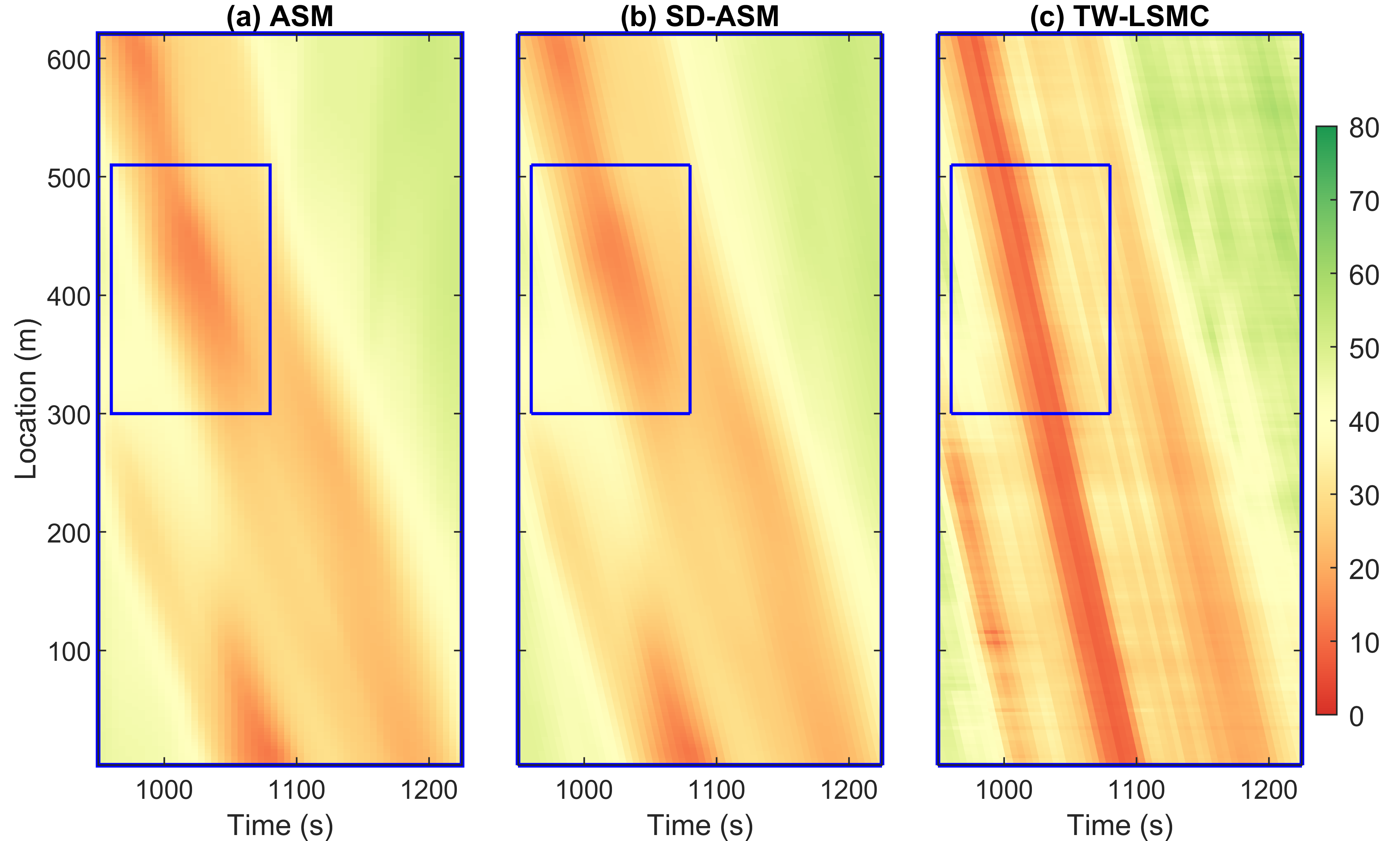}
  \captionsetup{font={small}}
  \caption{Comparison between rectangular and oblique grid-based methods (Zoom In).
  }
  \label{fig: Oblique grid}
  \end{center}
\end{figure}

The SD-ASM exhibits overall similar estimation effects to ASM, with their primary differences shown in local perspectives due to their utilization of different grid structures. Consequently, SD-ASM is not depicted in Fig. \ref{fig: Visualization-TSE}. Instead, Fig. \ref{fig: Oblique grid} zooms in on the nuanced differences between ASM and SD-ASM to more clearly illustrate the impact of employing oblique versus rectangular grids. The visualized period in Fig. \ref{fig: Oblique grid} is from the 950s to 1225s, corresponding to the left blue rectangle in Fig. \ref{fig: Visualization-TSE}. As depicted in Fig. \ref{fig: Oblique grid}(a), the ASM, which uses a rectangular grid, is prone to a noticeable aliasing effect, leading to speed discontinuities. In contrast, the SD-ASM, which adopts an oblique grid under the same spatiotemporal resolution, presents a significantly smoother profile, shown in Fig. \ref{fig: Oblique grid}(b). This enhancement in performance is further supported by reductions in the average RMSE/MAE and standard deviation as detailed in Tab. \ref{tab: TSE performance}, highlighting the effectiveness of the oblique grid in delivering consistent estimates and promoting state homogeneity. Furthermore, A direct comparison between Fig. \ref{fig: Oblique grid}(b) and Fig. \ref{fig: Oblique grid}(c) reveals that within the same oblique grid framework, the proposed low-rank-based TW-LSMC reconstructs more complete and precise shockwaves than the smoothing-based SD-ASM, showcasing the superior capability of TW-LSMC.

To comprehensively evaluate the model performance, we design multiple testing scenarios with varying levels of vehicle penetration rates. Specifically, we configure the penetration rates of connected vehicles (CVs) as 3$\%$, 5$\%$, 10$\%$, and 15$\%$, and repeat the experiment 20 times in each CV penetration scenario by randomly selecting different vehicle trajectories. Tab. \ref{tab: TSE performance} summarizes the RMSE (km/h) and MAE (km/h) with standard deviations of all models, numerically demonstrating their TSE performance. Overall, the proposed TW-LSMC outperforms the baseline models, particularly showing strength in scenarios with lower CV penetration rates, such as 3$\%$ and 5$\%$. In comparison to the ASM, the SD-ASM shows enhanced performance in terms of both accuracy and reduced variability across all scenarios, attributed to the integration of the oblique grid, which effectively addresses speed inconsistencies. Meanwhile, the PSM, augmented with a synchronized phase-based kernel, surpasses the ASM in penetration scenarios from 5$\%$ to 15$\%$. However, its performance dips below that of the ASM at the lowest penetration rate of 3$\%$, probably because more complex models usually require more data for training.

Notably, under the extremely sparse data environment of 3$\%$ CV penetration, the pure data-driven method STH-LRTC degrades sharply. This is because the Hankelization operation in STH-LRTC, which only integrates the limited observations from a surrounding orthogonal area of size $\tau_s \times \tau_t$, struggles to capture sufficient data under such extreme conditions. Conversely, the proposed TW-LSMC continues to provide accurate speed estimation results, offering more reliable support for refined proactive traffic control and management applications. Therefore, in the early stage of mixed conventional and connected environments with low CV penetration, the proposed TW-LSMC emerges as the more appropriate option.

\begin{table*}[t]
  \centering
  \caption{TSE performance comparison in average RMSE (km/h) and MAE (km/h) with the standard deviation.}
    \begin{threeparttable}
    \begin{tabular}{cccccrcccc}
    \toprule
    \multirow{2}[4]{*}{} & \multicolumn{4}{c}{RMSE (km/h)} &       & \multicolumn{4}{c}{MAE (km/h)} \\
\cmidrule{2-5}\cmidrule{7-10}          & CV-3\% & CV-5\% & CV-10\% & CV-15\% &       & CV-3\% & CV-5\% & CV-10\% & CV-15\% \\
\cmidrule{1-5}\cmidrule{7-10}     {LSMC \cite{candes2011robust}} &  {15.18 ± 0.94} &  {13.74 ± 0.72} &  {11.29 ± 0.67} &  {9.42 ± 0.60} &  {} &  {13.15 ± 0.71} &  {10.65 ± 0.61} &  {8.40 ± 0.49} &  {6.90 ± 0.40} \\
     {LWR-CG \cite{lu2023physics}} &  {10.75 ± 0.56} &  {8.52 ± 0.42} &  {6.91 ± 0.35} &  {6.55 ± 0.18} &  {} &  {7.54 ± 0.31} &  {6.32 ± 0.27} &  {5.45 ± 0.22} &  {4.91 ± 0.16} \\
    ASM \cite{treiber2011reconstructing}   & 9.89 ± 0.65 & 8.27 ± 0.51 & 6.86 ± 0.32 & 6.45 ± 0.19 &       & 7.27 ± 0.40 & 6.09 ± 0.36 & 5.12 ± 0.20 & 4.85 ± 0.14 \\
     {SD-ASM \cite{tsanakas2022generating}} &  {9.82 ± 0.29} &  {8.20 ± 0.20} &  {6.77 ± 0.10} &  {6.36 ± 0.03} &  {} &  {7.25 ± 0.15} &  {6.08 ± 0.11} &  {5.07 ± 0.04} &  {4.78 ± 0.02} \\
     {PSM \cite{rempe2017phase}} &  {10.85 ± 3.58} &  {8.08 ± 0.99} &  {6.63 ± 0.33} &  {6.34 ± 0.32} &  {} &  {7.44 ± 1.94} &  {5.93 ± 0.56} &  {4.99 ± 0.25} &  {4.73 ± 0.23} \\
    STH-LRTC \cite{wang2023low} & 36.06 ± 3.65 & 8.66 ± 2.84 & 5.89 ± 1.26 & \textbf{5.04 ± 0.32} &       & 23.6 ± 2.10 & 6.10 ± 1.37 & 4.37 ± 1.04 & \textbf{3.72 ± 0.18} \\
    TW-LSMC & \textbf{9.53 ± 0.75} & \textbf{7.56 ± 0.55} & \textbf{5.76 ± 0.44} & 5.14 ± 0.25  &       & \textbf{7.13 ± 0.45} & \textbf{5.66 ± 0.46} & \textbf{4.30 ± 0.24} & 3.86 ± 0.16 \\
    \bottomrule
    \end{tabular}%
    \begin{tablenotes}
      \item[] Delay-embedding lengths: $^{a}$  $\tau_s=60$,  $\tau_t=60$, $^{b}$  $\tau_s=40$,  $\tau_t=50$, $^{c}$ $\tau_s=30$, $\tau_t=50$,   $^{d}$ $\tau_s=20$, $\tau_t=50$.
    \end{tablenotes}
    \end{threeparttable}%
    \label{tab: TSE performance}%
\end{table*}%

\subsection{Robust Traffic State Estimation (RQ2)}
\label{subsec: performance evaluation: RTSE}

\begin{figure}[htbp]
  \begin{center}
  \includegraphics[width=3.5in]{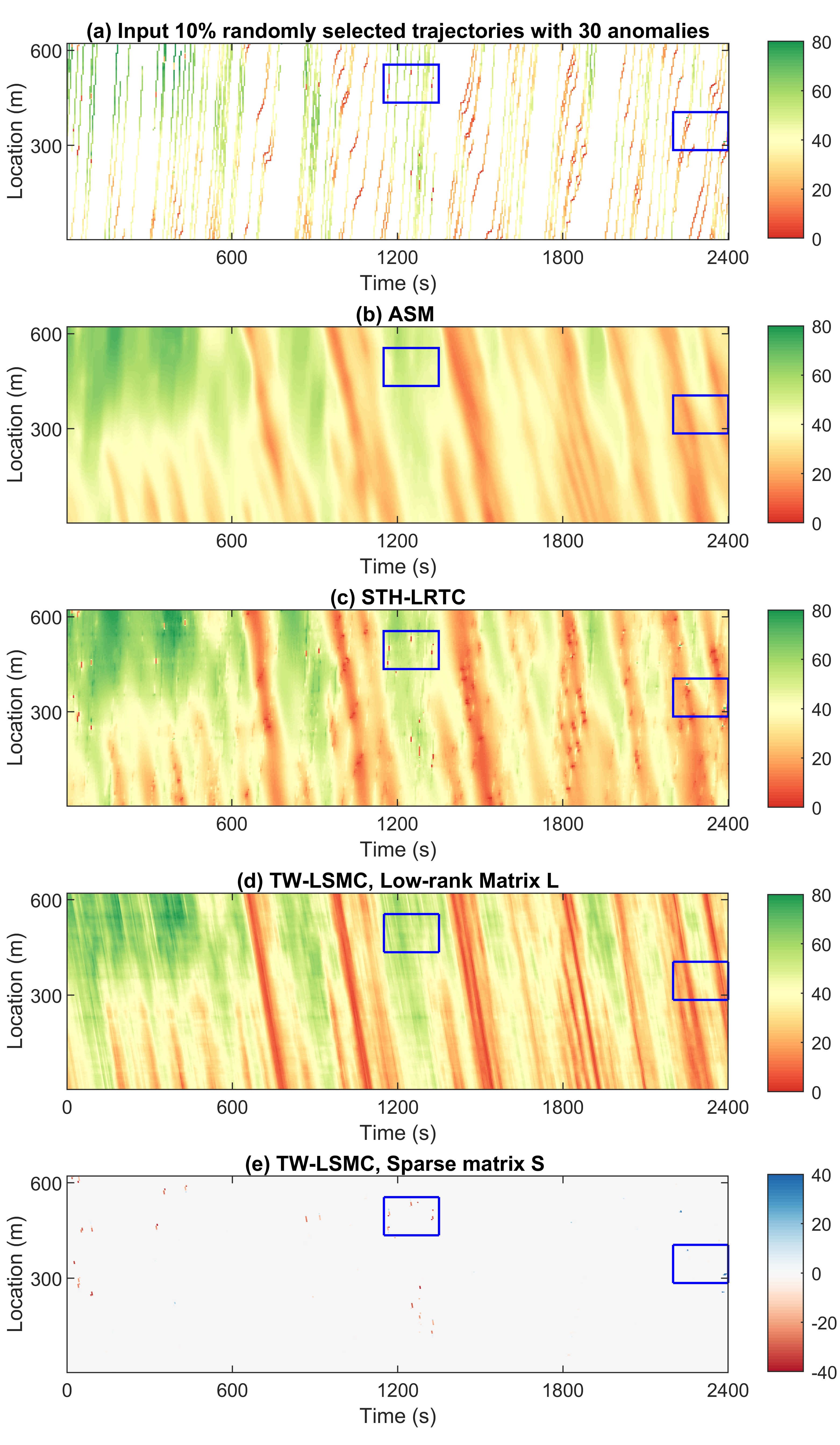}
  \captionsetup{font={small}}
  \caption{An RTSE experiment on the NGSIM dataset:  (a) The observed traffic speed; (b) The estimation result by ASM; (c) The estimation results by STH-LRTC; (d) The recovered low-rank traffic state matrix by the proposed TW-LSMC; (e) The recovered sparse anomaly matrix by the proposed TW-LSMC, positive entries (blue): type II data corruptions, negative entries (red): type I data corruptions.}
  \label{fig: Visualization-RTSE}
  \end{center}
\end{figure}

To directly showcase the robustness of TSE models, we initiate our evaluation with a certain data corruption scenario. Fig. \ref{fig: Visualization-RTSE} presents the estimation results of the proposed TW-LSMC alongside two state-of-the-art (SOTA) baseline methods. Given that the ASM, SD-ASM, and PSM methodologies are all based on smoothing techniques and that the PDEs utilized in the LWR-CG model have similar effects to the smoothing kernel, we choose ASM as the representative method to depict the robust TSE performance for this group. Fig. \ref{fig: Visualization-RTSE}(a) shows the observed traffic speed of 10$\%$ randomly selected trajectories with 30 type I and II data corruptions defined in Eq. \eqref{eq: type I corruption} and  Eq. \eqref{eq: type II corruption}. The ASM's estimations, depicted in Fig. \ref{fig: Visualization-RTSE}(b), show a relative insensitivity to corruption, attributed to the anomaly-mitigating effect of its weighted smoothing operation. When compared to ASM, the STH-LRTC method yields more accurate results in areas unaffected by corruption. However, its performance significantly declines within corrupted zones (see the blue rectangles in Fig. \ref{fig: Visualization-RTSE}(c)), owing to the presumption of uncorrupted speed observations in the Hankel tensor construction. Fig. \ref{fig: Visualization-RTSE}(d) and (e) show the TW-LSMC's reconstructed low-rank traffic state matrix and the sparse anomaly matrix respectively. The positive and negative values in Fig. \ref{fig: Visualization-RTSE}(e) refer to the type II and I data corruptions, respectively. The low-rank matrix accurately provides complete structural traffic states, while the sparse matrix successfully detects both types of randomly injected corruptions (see the blue rectangles  Fig. \ref{fig: Visualization-RTSE}(e)), confirming the necessity of individually modeling the potential anomalies in a robust TSE model.

\begin{figure}[t]
  \begin{center}
  \includegraphics[width=2.5in]{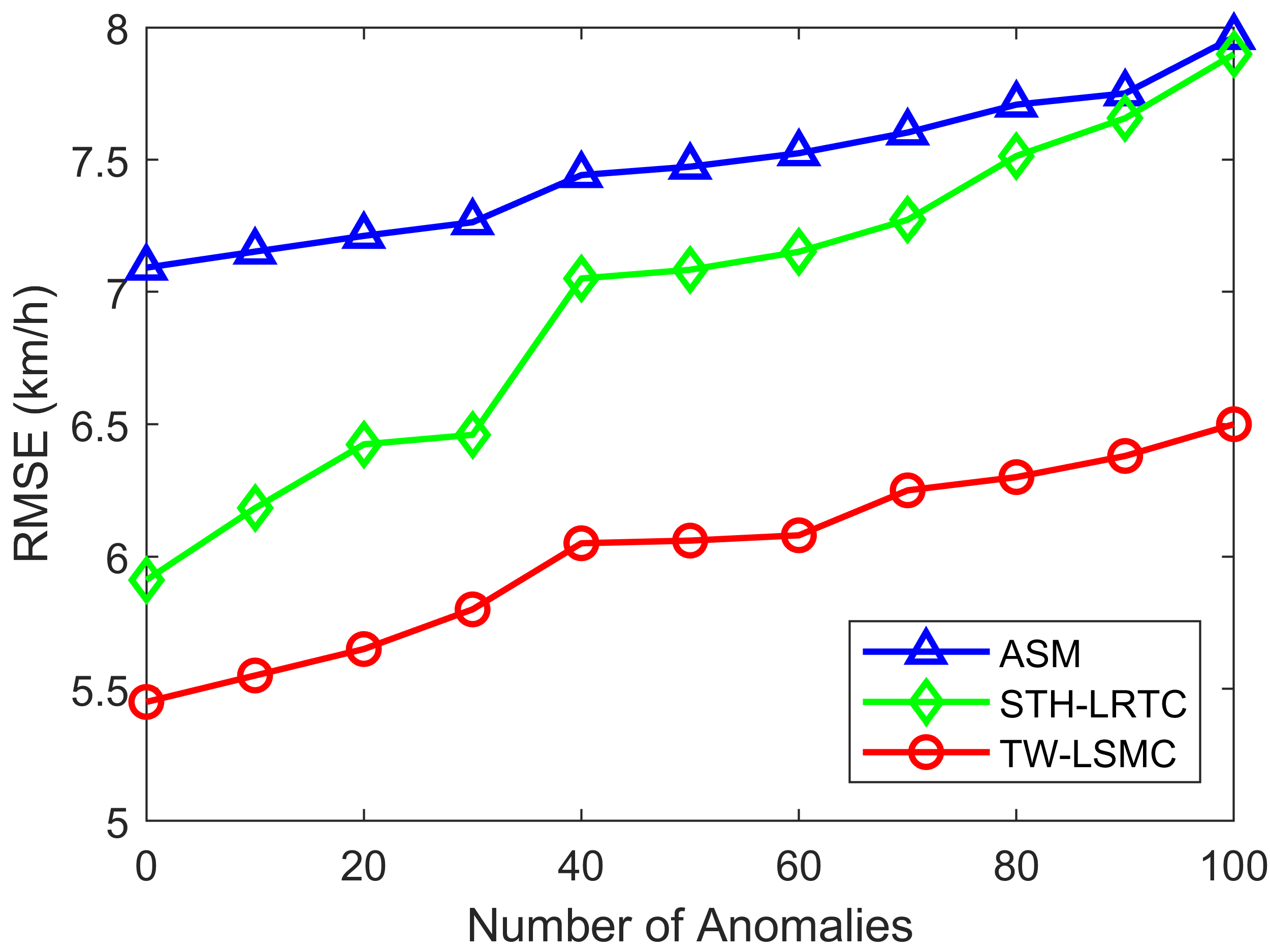}
  \captionsetup{font={small}}
  \caption{Robust TSE performance comparison under varying data corruption levels. For each corruption level, the number of type I and type II data anomalies/corruptions is the same.
  }
  \label{fig: RTSE performance comparison}
  \end{center}
\end{figure}

To comprehensively evaluate the model performance of robust traffic state estimation (RTSE) under varying data corruption levels, we randomly inject a variety number of type I and type II data corruptions into observations. 
Fig. \ref{fig: RTSE performance comparison} displays the performance (in RMSE) of the proposed and alternative methods. As the corruption level increases, our TW-LSMC model which leverages a low-rank and sparse representation exhibits remarkable robustness with RMSE values rising modestly from 5.5 to 6.5 km/h. In contrast, the performance of the low-rank Hankel tensor-based STH-LRTC method deteriorated significantly, with RMSE increasing from 6.0 to 8.0 km/h. These results demonstrate the effectiveness of the anomaly-tolerant module in the proposed method.
In the meantime, ASM performs insensitivity to the changes in corruption level, with RMSE increasing from 7.0 to 8.0 km/h, because the smoothing operation in ASM can mitigate the negative effect of anomalies to a certain extent. The inadequate performance of ASM mainly stems from the basic estimation ability in anomaly-free scenarios, which is due to the limitation of smoothing operation's capability to capture traffic state dependencies, as previously discussed in subsection \ref{subsec: performance evaluation: TSE}.

\subsection{Sensitivity analysis (RQ3)}
\label{subsec: Sensitivity analysis}

\begin{figure}[t]
  \begin{center}
\includegraphics[width=2.5in]{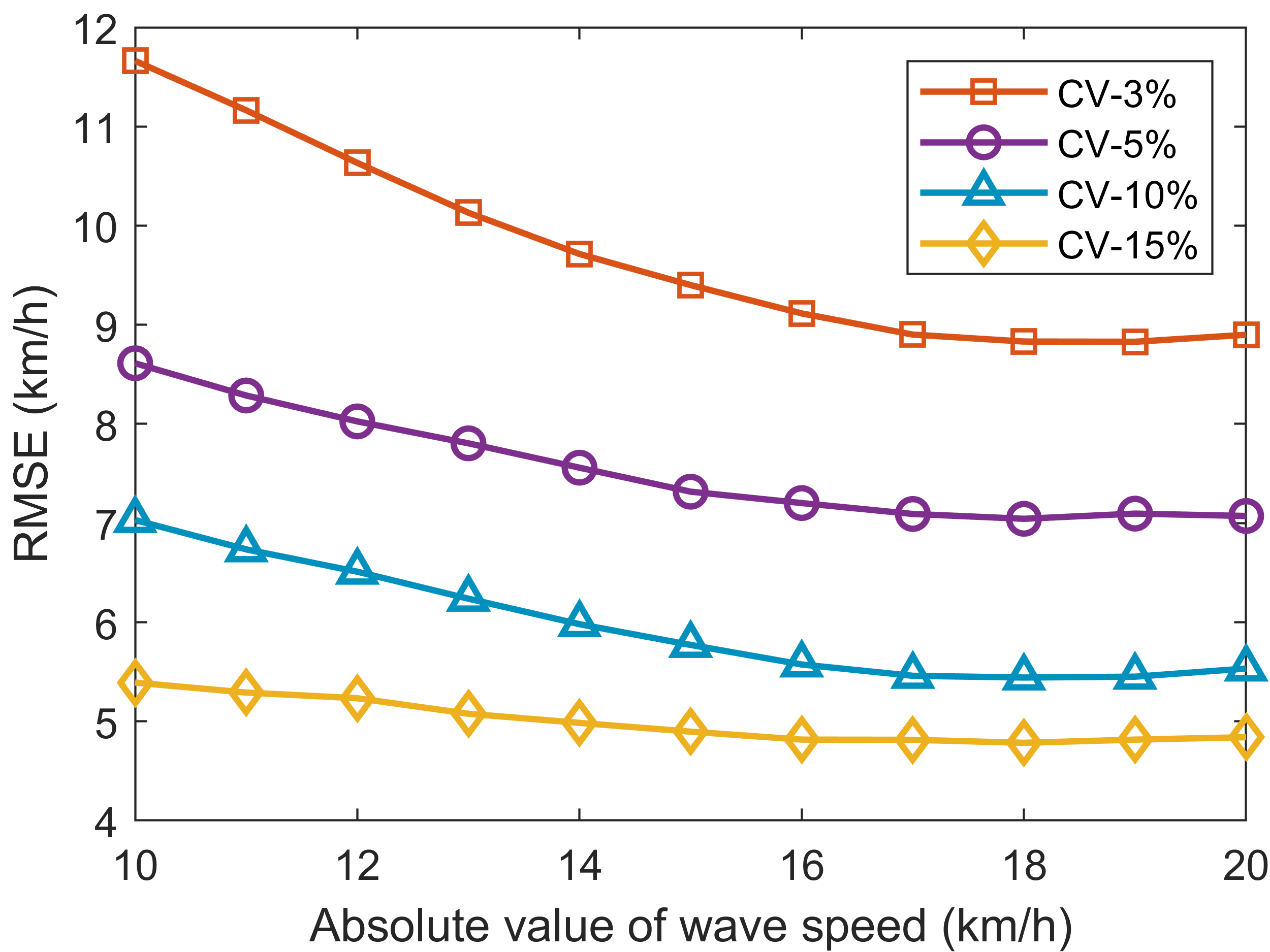}
\caption{Sensitivity of the wave speed on NGSIM dataset.}
\label{fig: sensitivity}
  \end{center}
\end{figure}

The backward wave speed is an important parameter in the proposed TW-LSMC, as it introduces valuable physical traffic propagation knowledge. The significance of traffic wave prior is tested in the ablation study (subsection \ref{subsec: Ablation study}).
We investigate the wave speed sensitivity of the proposed method under different CV penetrations using the NGSIM dataset and summarize the model performance (in RMSE) in Fig. \ref{fig: sensitivity}.
The backward wave speeds around the world generally range from -10 to -20 km/h \cite{chen2014periodicity, he2019constructing}. 
For three CV penetration scenarios, the TW-LSMC achieved the best performance with the lowest RMSE when the wave speed equals -18 km/h, indicating the actual backward wave speed of the NGSIM dataset, which is consistent with the estimated value in \cite{chen2022integrated, chen2014estimating}.
It is also interesting to find that the performance of the TW-LSMC reaches a stable platform when the absolute value of the wave speed parameter is larger than 16 km/h, suggesting a recommended wave speed value range that provides acceptable performance.  It is another indication of the robustness of the proposed method, as a stable high-performance interval of the core model parameter is useful in practical scenarios.

\subsection{Ablation study (RQ4)}
\label{subsec: Ablation study}

To inspect the significance of each component of the proposed method, we conduct an ablation study to compare the performance of model variation by repeating the experiments 20 times on the RTSE scenarios using 10$\%$  trajectories with 30 type I and type II data corruptions. We examine three variations of the proposed method:  (1) In TW-LSMC w/o TW, we adopt the conventional matrix construction instead of the one using the traffic wave prior. (2) In TW-LSMC w/o nonconvex, we replace the nonconvex truncated nuclear norm (TNN) function with the convex nuclear norm (NN) function defined in the subsection \ref{subsec: Notations}. (3) In TW-LSMC w/o S term, we remove the sparse anomaly components in TW-LSMC, and only the low-rank matrix is preserved. Results of different variations are shown in Tab. \ref{tab: Ablation studies}.

\begin{table}[t]
  \centering
  \caption{Average RMSE (km/h) and MAE (km/h) with standard deviation of variant methods in ablation study.}
  \scalebox{0.91}{
    \begin{tabular}{ccccc}
      \toprule
      \multirow{2}[4]{*}{Metric} & \multicolumn{4}{c}{Method} \\
  \cmidrule{2-5}          & TW-LSMC & w/o TW (LSMC) & w/o nonconvex   &  w/o S term \\
      \midrule
    RMSE  & 6.07 $\pm$ 0.43 & 11.40 $\pm$ 0.51 & 7.13 $\pm$ 0.86  & 6.64 $\pm$ 0.41 \\
    MAE   & 4.47 $\pm$ 0.24 & 8.50 $\pm$ 0.38 & 5.13 $\pm$ 0.45  & 4.80 $\pm$ 0.25 \\
    \bottomrule
    \end{tabular}}%
  \label{tab: Ablation studies}%
\end{table}%

From the results, we can observe that the performance of all the variations degraded, indicating that each component contributes to the overall improvement of TW-LSMC remarkably. After replacing the nonconvex TNN term, the errors show an obvious increase, demonstrating that the nonconvex rank surrogate function is more capable of capturing the traffic state's low-rank nature (i.e., spatiotemporal dependencies) than the convex one. 
It is notable that without traffic wave prior, the accuracy decreases sharply. This indicates that the vanilla low-rank matrix completion method is incapable of capturing the traffic dynamics and propagation characteristics. The ablation studies on sparse matrix  $\mathbf{S}$ term manifest that potentially corrupted data should be considered and modeled in the TSE model, and the increments of RMSE and MAE verify this finding.

\subsection{Computation Performance (RQ5)}
\label{subsec: Computation Performance}

To demonstrate the computational performance of the proposed TW-LSMC model, we first theoretically analyze the temporal computational complexity of the proposed and four representative baseline methods in Tab. \ref{tab: computational complexity}, and then give empirical running time evidence in Tab. \ref{tab: running time}.

\subsubsection{Computational complexity comparison}

Before analyzing the computational complexity, we denote the spatial and temporal length of the input traffic state matrix by $L$ and $T$ and the number of iterations by $k$. 
The most time-consuming step when training LWR-CG is within the shared layer that connects the input layer with subsequent layers, which contributes to a complexity  $\mathcal{O} \left( kbLTM \right)$, where $M$ is the number of hidden neurons, $b$ is the batch size, and $k$ is the number of epochs.
The complexity of ASM is dominated by the calculation of two free-flow and congested speed fields. For each single spatiotemporal location $\left( l,t \right) $, all the data observations are used for the calculation. Thus, the complexity of ASM is $\mathcal{O} \left( NLT \right) $, where $N$ is the number of observations.
As a result, the computation time of ASM will dramatically increase when more high-resolution data are used. The SD-ASM shares the same complexity with ASM.
The most complex step in PSM is the convolution process when calculating phase-dependent speeds, contributing to a complexity  $\mathcal{O} \left( LT\tau\sigma \right)$, where $\tau$ and $\sigma$ are the temporal and spatial lengths of the convolution kernel.
For the STH-LRTC, the most complex step is the update of the Hankel tensor $\mathcal{X} \in \mathbb{R} ^{\tau _s\times \tau _t\times \left( L-\tau _s+1 \right) \times \left( T-\tau _t+1 \right)}$, where $\tau_s$ and $\tau_t$ are the embedding lengths of Hankel tensor.
Specifically, it applies one SVD on the reshaped matrix $\mathcal{X} _{\Box}\in \mathbb{R} ^{\tau _s\tau _t\times \left( L-\tau _s+1 \right) \left( T-\tau _t+1 \right)}$, contributing to a per-iteration computational complexity of $\mathcal{O} \left( \tau _{s}^{2}\tau _{t}^{2}\left( L-\tau _s+1 \right) \left( T-\tau _t+1 \right) \right) $. Therefore, the computational complexity of STH-LRTC will increase when larger embedding lengths are configured in the Hankel tensor. The computational complexity of the proposed method is analyzed in subsection \ref{subsec: Computational complexity}.

In our experiments, the spatiotemporal size of the input traffic state matrix $L$ and $T$, the number of required convergence iterations $k$, the number of data observations $N$, the embedding lengths $\tau_t$ and $\tau_s$, and the kernel lengths $\tau$ and $\sigma$ are noted at the bottom of Tab. \ref{tab: computational complexity}.
To ensure an identical spatiotemporal reconstruction area, the temporal size of the oblique grid-based input TSM is slightly larger than the orthogonal grid-based TSM, i.e., $505$ and $480$.
As the spatial length $L$ is much smaller than the number of observations $N$ in ASM and the squared embedding length $\tau _{s}^{2}\tau _{t}^{2}$ in STH-LRTC, we theoretically prove that our method is more computationally efficient than the state-of-the-art (SOTA) data-driven methods.

\begin{table}[t]
  \centering
  \caption{The computational complexity of the baseline and proposed models. }
  \begin{threeparttable}
    \begin{tabular}{cc}
    \toprule
      Method    & Computational complexity \\
    \midrule
    LWR-CG &  $\mathcal{O} \left( kbLTM \right)$ $^{a}$ \\
    ASM/SD-ASM   &  $\mathcal{O} \left( NLT \right) $$^{b}$ \\
    PSM & $\mathcal{O} \left( LT\tau\sigma \right)$ $^{c}$ \\
    STH-LRTC & $\mathcal{O} \left( k\tau _{s}^{2}\tau _{t}^{2}\left( L-\tau _s+1 \right) \left( T-\tau _t+1 \right) \right) $ $^{d}$  \\
    TW-LSMC &  $\mathcal{O} \left( kL^2T \right) $$^{e}$ \\
    \bottomrule
    \end{tabular}%
    \begin{tablenotes}
      \item[a]  $k=10000$, $L=207$, $T=480$, $b = N/3, M=125 $.
      \item[b]  $L=207$, $T=480$ (ASM), $T=505$ (SD-ASM), $N= 2980 \sim 14904 $.
      \item[c]  $L=207$, $T=480$, $\tau = 20 \sim 500, \sigma = 100 \sim 1000 $.
      \item[d]  $k=80\sim100$, $L=207$, $T=480$, $\tau_s = 20\sim60$, $\tau_t = 50$.
      \item[e]  $k=50\sim100$, $L=207$, $T=505$.
  \end{tablenotes}
\end{threeparttable}%
  \label{tab: computational complexity}%
\end{table}%

\subsubsection{Running time comparison}
To further compare the computational efficiency of TSE models, we summarize the running time of the proposed and baseline models under various CV penetrations in Tab. \ref{tab: running time}.
Overall, the total running time of the proposed TW-LSMC consistently outperforms the SOTA models in all scenarios, demonstrating its promising and reliable computational ability regardless of missing data characteristics. Among all the models evaluated, the LWR-CG requires the longest training time, which can be attributed to its extensive number of iterations and batch size.
The running time of ASM dramatically grows with the increase of the CV penetration since the computational complexity of ASM is positively related to the amount of data.
In contrast, the running time of STH-LRTC decreases, because the Hankel tensor can be smaller when input data are more sufficient, e.g, $\tau_s =60$ m is used in the CV-3$\%$ case and $\tau_s = 20$ m is applied in the CV-15$\%$ case. 
The running time of PSM is smaller than ASM because only the speed observations within convolution kernels are used in PSM when estimating phase-dependent speeds, instead of using all observations in ASM.
In summary, the theoretical analysis and empirical evidence both confirm the computational superiority of the proposed method.

\begin{table}[t]
  \centering
  \caption{The average running time (s) with the standard deviation in NSGIM data experiment under various CV penetrations.}
  \resizebox{0.5\textwidth}{!}{
    \begin{tabular}{cccccc}
    \toprule
    \multirow{2}[4]{*}{Scenarios} & \multicolumn{5}{c}{Method} \\
\cmidrule{2-6}          &  {LWR-CG} & ASM   &  {PSM} & STH-LRTC & TW-LSMC \\
    \midrule
    CV-3\% &  {6430.2 ± 70.3} & 24.8 ± 2.3 &  {24.8 ± 1.3} & 1997.3 ± 162.3 & \textbf{0.85 ± 0.06} \\
    CV-5\% &  {8742.6 ± 83.7} & 42.9 ± 3.4 &  {28.1 ± 1.5} & 524  ± 19.0 & \textbf{0.81 ± 0.02} \\
    CV-10\% &  {11384.1 ± 96.1} & 84.6 ± 4.4 &  {42.1 ± 2.0} & 337.8 ± 3.1 & \textbf{0.78 ± 0.02} \\
    CV-15\% &  {13215.8 ± 103.0} & 122.7 ± 4.5 &  {55.7 ± 2.7} & 185.7 ± 1.6 & \textbf{0.77 ± 0.06} \\
    \bottomrule
    \end{tabular}}%
  \label{tab: running time}%
\end{table}%

\section{Discussion} \label{sec: discussions}

\subsection{Use conditions}
The present work proposes a traffic wave-based low-rank and sparse matrix completion model that utilizes trajectory data obtained from connected vehicles (CVs). The proposed model is also compatible with fixed detector data, as any observations can be transformed into the input entries of the traffic state matrix to improve the state estimation accuracy. The most important aspects of this study are: (1) No physical traffic model is used, only a backward traffic wave speed is required, which generally ranges from -10 km/h to -20 km/h around the world \cite{chen2014periodicity, he2019constructing} and the model performance is robust to this parameter selection when the absolute wave speed parameter larger than 16km/h; (2) No data pre-processing procedures are required, and the model can accommodate corrupted input data; (3) No extensive historical data are required for model training, i.e., the model is unsupervised; (4) Only small penetration rates, e.g., 5$\%$ CV deployed in the freeway, are sufficient to provide traffic speed estimations with small errors.

\subsection{Limitations}
As discussed, the applicability of the current research is straightforward. One limitation of this study is only traffic speed states are being estimated. The direct application of the proposed methodology employing CV trajectories for traffic density estimation may be biased. This is due to the deviated traffic density observations measured from the CVs, compounded by the absence of additional physical models (e.g., fundamental diagram). A viable strategy to resolve it involves the integration of connected and automated vehicles (CAVs) that allow the collection of space or time headway from surrounding vehicles, thereby enabling the generation of unbiased traffic density measurements within small spatiotemporal grid cells, exemplified by cells of 3 meters by 5 seconds.

\section{Conclusion} \label{sec: conclusions}

In this study, we propose a simple and efficient matrix completion model for traffic state estimation (TSE) using sparse vehicle trajectory data. 
Inspired by the traffic wave prior, we construct the traffic state matrix with oblique grids to capture the recurrent traffic dynamics and directional traffic propagation characteristics. To enhance the robustness of the proposed TSE model, we design an anomaly-tolerant module to detect and remove anomalies in traffic state observations.
Extensive experiments indicate that (1) the oblique grid-based modeling is able to capture traffic dynamics and achieves reliable estimation performance, especially in extremely sparse data conditions, (2) the model consistently performs robustness to various data corruption levels, and (3) the model is robust to wave speed parameters, can adapt to diverse traffic scenarios and is more computationally efficient than the SOTA data-driven methods.

There are several further directions for future study. First, the present model is designed for the traffic speed estimation problem and could be extended to estimate volume, density, and other traffic state variables. Second, while this study addresses random non-Gaussian data corruption (C2), future investigations could explore more intentional corruption such as cyber-attacks. Third, the proposed method is evaluated using a connected vehicle (CV) trajectory dataset. Future endeavors could look into applying this methodology to multi-source traffic datasets or extended floating car data (xFCD) from connected and automated vehicles (CAVs).

\ifCLASSOPTIONcaptionsoff
  \newpage
\fi




\bibliographystyle{IEEEtranN}
\bibliography{ref}

\vspace{-2em}

\begin{IEEEbiography}[{\includegraphics[width=1in,height=1.25in,clip,keepaspectratio]{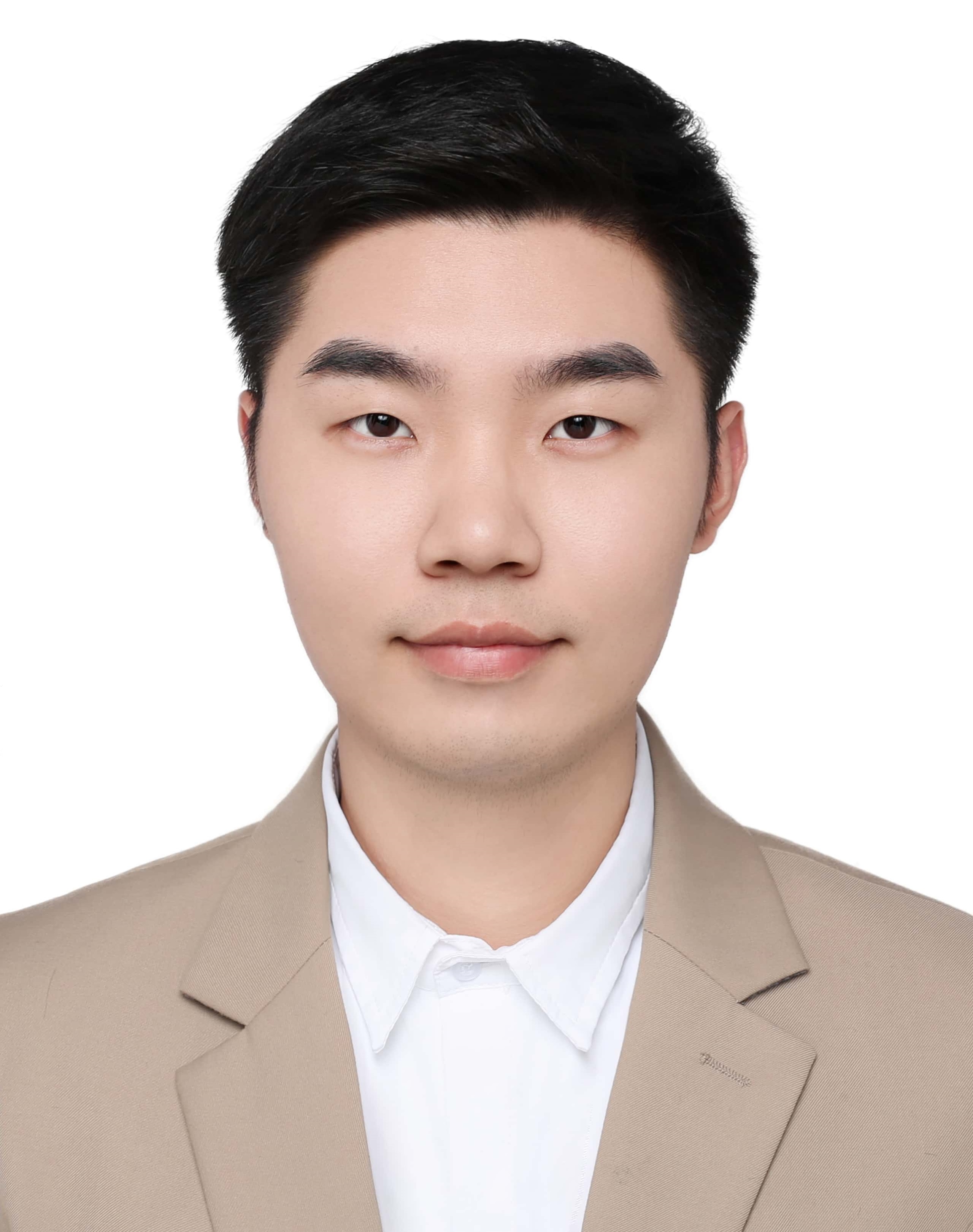}}]{Yang He}
    received his B.E. degree in Traffic Engineering from Chang’ an University, Xi'an, China, in 2020. He is currently pursuing the Ph.D. degree in the Intelligent Transportation System Research Center at Southeast University. His current research interests include traffic state estimation, network modeling, and low-rank modeling.
\end{IEEEbiography}

\vspace{-2em}

\begin{IEEEbiography}[{\includegraphics[width=1in,height=1.25in,clip,keepaspectratio]{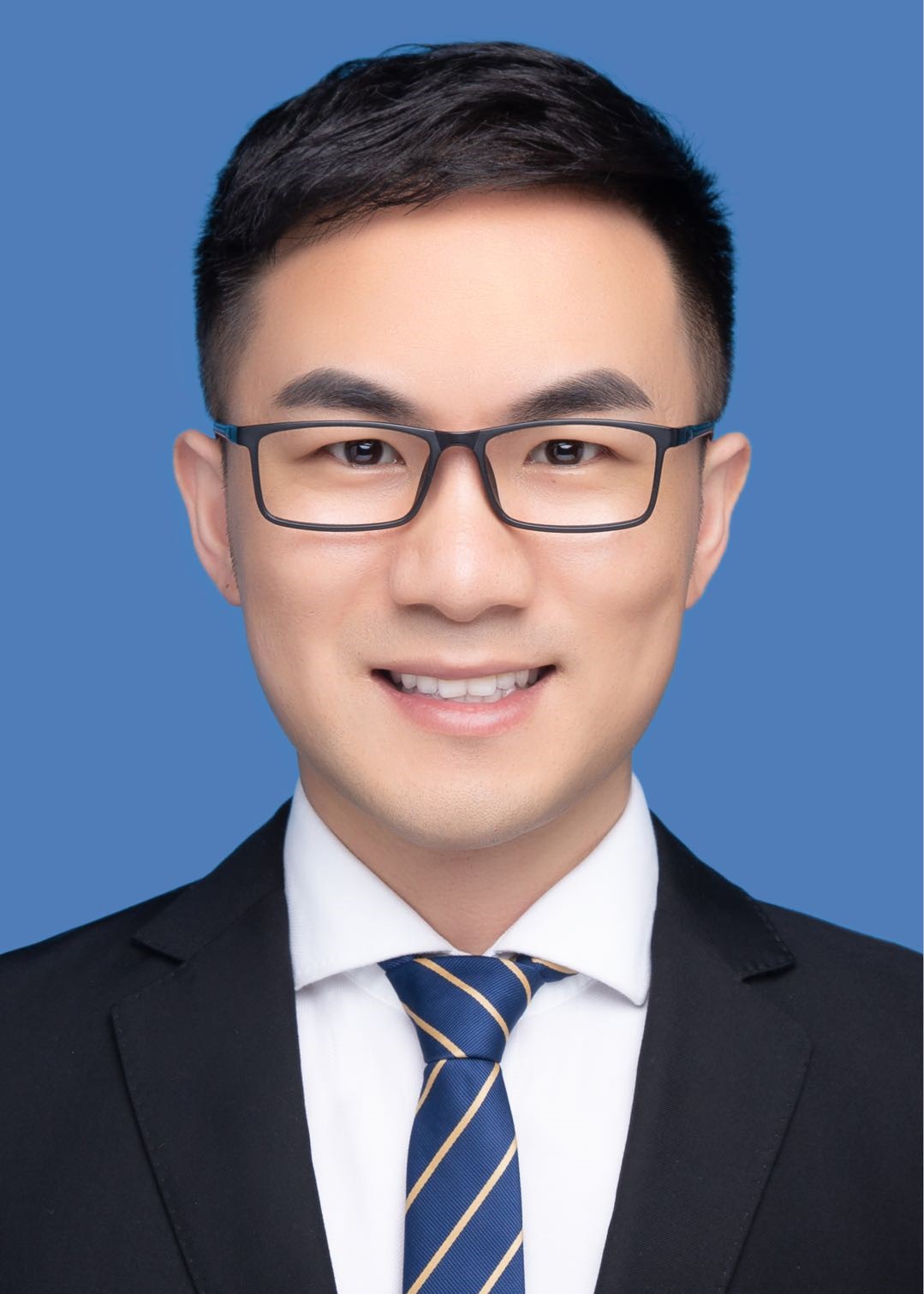}}]{Chengchuan An}
    received his Ph.D. degree in Transportation Engineering from Southeast University, Nanjing, China, in 2019. From 2014 to 2016, he was a visiting scholar at the Department of Civil and Architectural Engineering and Mechanics, University of Arizona, USA. Since 2020, he has been a post-doctor in the Intelligent Transportation System Research Center at Southeast University. His current research interests include intelligent traffic signal control systems and traffic data mining.
\end{IEEEbiography}

\vspace{-2em}

\begin{IEEEbiography}[{\includegraphics[width=1in,height=1.25in,clip,keepaspectratio]{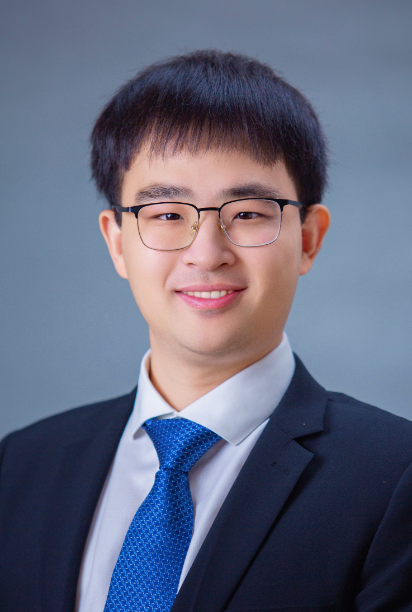}}]{Yuheng Jia}
    received the B.S. degree in automation
    and the M.S. degree in control theory and engineering
    from Zhengzhou University, Zhengzhou, China,
    in 2012 and 2015, respectively, and the Ph.D. degree
    in computer science from the City University of
    Hong Kong, SAR, China, in 2019.
    He is currently an associate professor with the
    School of Computer Science and Engineering,
    Southeast University, China. His research interests
    include machine learning, Bayesian method, spectral
    clustering and low-rank modeling.
\end{IEEEbiography}

\vspace{-2em}

\begin{IEEEbiography}[{\includegraphics[width=1in,height=1.25in,clip,keepaspectratio]{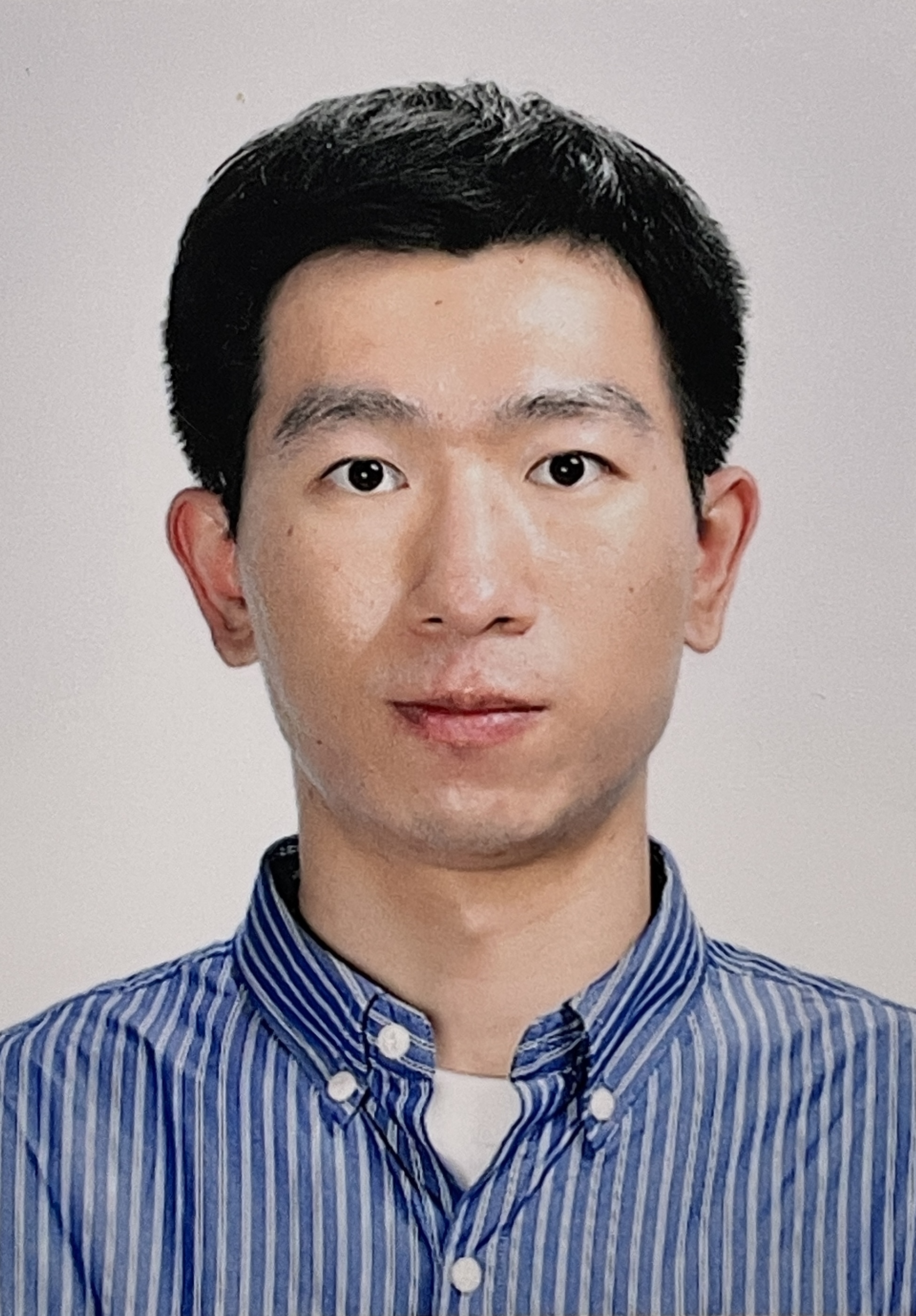}}]{Jiachao Liu}
  received his B.S. degree in Traffic Engineering from Dalian University of Technology, Dalian, China, in 2017, M.S. Degree in Transportation Engineering from Southeast University, Nanjing, China, in 2020, and M.S. Degree in Machine Learning from Carnegie Mellon University, USA, in 2023. He is currently pursuing the Ph.D. degree in Civil and Environmental Engineering at Carnegie Mellon University, USA. His research interests include transportation network modeling, simulation, optimization and machine learning.
\end{IEEEbiography}

\vspace{-2em}

\begin{IEEEbiography}[{\includegraphics[width=1in,height=1.25in,clip,keepaspectratio]{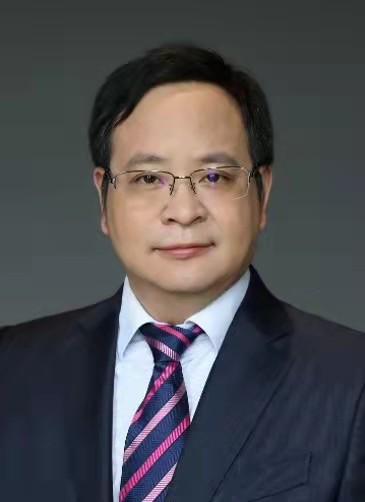}}]{Zhenbo Lu}
    received the Ph.D. degree in Traffic Information Engineering and Control from Southeast University, Nanjing, China, in 2011. He is currently an Associate Professor with the Intelligent Transportation System Research Center, Southeast University. His main research interests include transportation planning, traffic simulation, and intelligent transportation systems.
\end{IEEEbiography}

\vspace{-2em}

\begin{IEEEbiography}[{\includegraphics[width=1in,height=1.25in,clip,keepaspectratio]{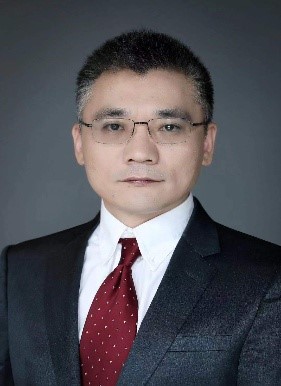}}]{Jingxin Xia}
is a professor at the Intelligent Transportation System Research Center, Southeast University, Nanjing, China. He received the Ph.D. degree in Transportation Engineering from the University of Kentucky, USA in 2006. He has published more than forty peer-reviewed papers so far, and his main research interests include traffic flow theory, transportation network modeling, traffic signal control, and intelligent transportation systems.
\end{IEEEbiography}

\end{document}